\documentclass[prc,showpacs,preprint,superscriptaddress]{revtex4}
\usepackage{graphicx}
\usepackage{dcolumn}
\usepackage{bm}

\begin{document}

\title{Microscopic description of quadrupole-octupole coupling
in Sm and  Gd  isotopes with the Gogny Energy Density Functional.}

\author{R. Rodr\'{\i}guez-Guzm\'an} 
\affiliation{Department of Chemistry, Rice University, Houston, Texas 77005, USA}
\affiliation{Department of Physics and Astronomy, Rice University, Houston, Texas 77005, USA}

\author{L. M. Robledo} 
\affiliation{Departamento  de F\'{\i}sica Te\'orica,
Universidad Aut\'onoma de Madrid, E-28049 Madrid, Spain}

\author{P. Sarriguren}
\affiliation{Instituto de Estructura de la Materia, IEM-CSIC, Serrano 123, E-28006 Madrid, Spain}

\date{\today}

\begin{abstract}
The interplay between the collective dynamics of
the quadrupole and octupole deformation degree of freedom
is discussed in a series of Sm and Gd isotopes both at 
the mean field level and beyond, including parity symmetry
restoration and configuration mixing. Physical
properties like negative parity excitation energies, E1 and E3 transition 
probabilities are discussed and compared to experimental data. Other relevant 
intrinsic quantities like dipole moments, ground state quadrupole moments or
correlation energies associated to symmetry restoration and configuration mixing 
are discussed. For the considered isotopes, the quadrupole-octupole
coupling is found to be weak and most of the properties of negative parity states can be
described in terms of the octupole degree of freedom alone.
\end{abstract}


\maketitle


\section{Introduction.}

The nuclear mass region with proton number $Z \approx 60$ and neutron number $N
\approx 90$ is receiving at present much attention, both experimental and
theoretically, since it is a region where nuclear structure collective effects
of different nature overlap \cite {butler_96}. Particularly interesting in this
context is the interplay between quadrupole transitional properties in $N
\approx 90$ isotones and octupole deformation manifestations in nuclei with
proton $Z \approx 56$ and neutron $N \approx 88$ numbers. On one hand, isotones
with $N \approx 90$ have been found as empirical realizations \cite{exp_x5} of
the critical point symmetry $X(5)$, introduced \cite{iachello} to describe
analytically the first order phase transition from spherical $[U(5)]$ to well
deformed $[SU(3)]$ nuclei. Such critical point symmetries, have recently been
studied within various microscopic approaches, either relativistic or
non-relativistic (see, for example, \cite
{ring_x5,ours_x5_1,ours_x5_2,Egido-Tomas-Nd} and references therein).

On the other hand, it is well known \cite{butler_96} that there is a 
tendency towards octupolarity around particular neutron/proton 
numbers, namely $N/Z=$ 34, 56, 88 and 134. The emergence of  
octupolarity  in these nuclear systems can be traced back to the 
structure of the corresponding single-particle spectra which exhibit 
maximum coupling between states of opposite parity, where the  
$(N+1,l+3,j+3)$ intruder orbitals interact with the $(N,l,j)$ 
normal-parity states through the octupole component of the effective 
nuclear Hamiltonian. When the mixing is strong enough, the nucleus 
displays an octupole deformed ground state \cite{butler_96}. In 
particular, for nuclei with $Z \approx  56$ ($N  \approx 88$) the 
coupling between the proton (neutron) single-particle states 
$h_{11/2}$ ($i_{13/2}$) and $d_{5/2}$ ($f_{7/2}$) has been 
considered as mainly responsible for mean field ground state 
octupolarity.

\begin{figure*}
\includegraphics[width=0.508\textwidth]{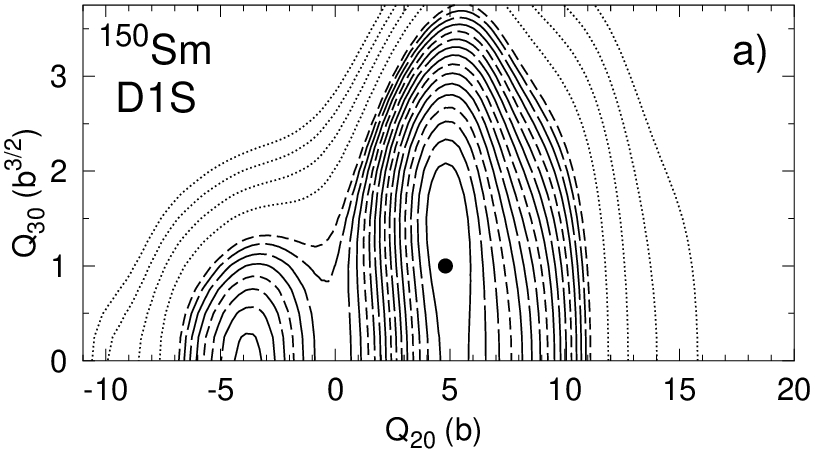}

\includegraphics[width=0.480\textwidth]{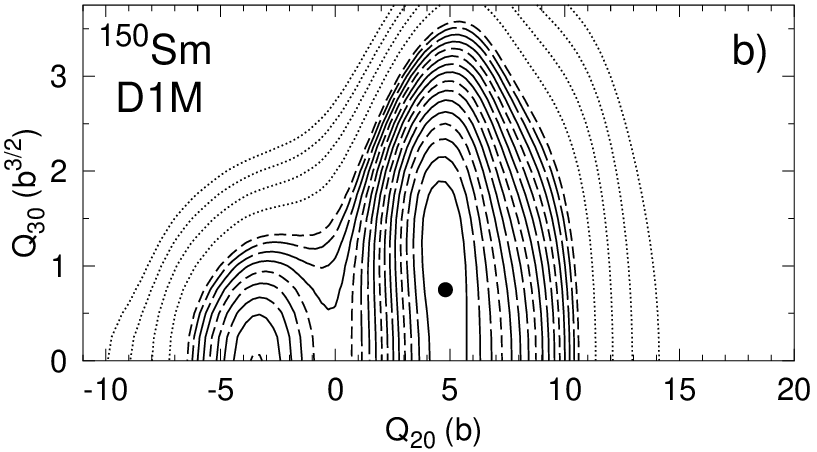} \par
\caption{MFPESs computed with the Gogny-D1S EDF in panel a) and Gogny-D1M
EDF in panel b) for the nucleus $^{150}$Sm. Taking the lowest energy
as a reference, the contour lines 
extend from 0.25 MeV  up to 4 MeV in steps of 0.25 MeV. 
Full, long dashed and short dashed contours are used successively 
to help identify contours more easily. Dotted lines 
correspond to contours starting at 5 MeV and extending to 8 MeV
in steps of 1 MeV.
}
\label{fig_mfq2q3_Sm}
\end{figure*}

The search for signatures of stable octupole deformations in atomic 
nuclei has been actively pursued during the last decades \cite
{butler_96,aberg_90}. As a main feature, octupole deformed even-even 
nuclei display particularly low-lying negative-parity $1^-$ states. 
In the case of stable octupole deformations, the $0^+$ and $1^-$ 
states represent the members of parity doublets, giving rise to 
alternating-parity rotational bands with enhanced $E1$ transitions 
among them. These fingerprints of   octupole deformations have 
already been found in the particular regions mentioned above, but 
especially in the rare-earth and actinide regions 
\cite{butler_96,aberg_90}.

For the  sample of nuclei considered in the present study (i.e., 
$^{146-154}$Sm and $^{148-156}$Gd), experimental fingerprints  have 
been obtained through the observation of  octupole correlations  at 
medium spins, as well as the crossing of the octupole and the ground state 
band, pointing to the fact that reflection symmetric and asymmetric 
structures coexist in $^{150}$Sm \cite{urban_87} and $^{148}$Sm \cite
{urban_91}. A recent study \cite{garrett_09} has analyzed the 
lowest four negative-parity bands in $^{152}$Sm and has found 
an emerging pattern of repeating excitations, built on the $0^+_2$ 
level and similar to that of the ground state, suggesting a complex 
shape coexistence in $^{152}$Sm.

The experimental findings \cite{urban_87,urban_91,garrett_09} 
mentioned above, already suggest  that it is timely and necessary to 
carry out systematic studies of the quadrupole-octupole  interplay 
in this and other regions of the nuclear chart, starting from modern 
(global) relativistic \cite{PR-VE-RI-2005,review-Bender} and/or 
non-relativistic \cite{review-Bender,gogny,gogny-d1s,bal08} 
nuclear Energy Density Functionals (EDFs),  with reasonable 
predictive power all over the nuclear chart.
  
Let us remark that the microscopic study of the dynamical (i.e., 
beyond mean field) quadrupole-octupole coupling in  the considered 
Sm and  Gd  isotopes  is also required to better understand the 
extent to which a picture of independent quadrupole and octupole 
excitations persists or breaks down  for nuclei with neutron number 
$N \approx 88$. This, together with the available  experimental 
fingerprints \cite{urban_87,urban_91,garrett_09} for octupolarity in 
the region, is one of the main reasons driving our choice of the 
nuclei $^{146-154}$Sm and $^{148-156}$Gd  as a representative sample 
to test the performance of the different approximations  and EDFs 
considered in the present study.

\begin{table} 
\label{tableSm}
\caption{Proton $E_{p}(Z)$ (MeV) and neutron $E_{p}(N)$ (MeV) pairing energies, dipole
$D_{0}$ (efm) moment, quadrupole $Q_{20}$(b) and octupole $Q_{30}$(b$^{3/2}$) 
moments at the minima of the MFPESs for the isotopes $^{146-154}$Sm. Results are given for both Gogny-D1S and 
Gogny-D1M EDFs. 
}
\begin{tabular}{cccccccccccccc}
\hline
Nucleus  && $E_{p}(Z)$ & $E_{p}(N)$ & $D_{0}$  & $Q_{20}$ & $Q_{30}$ &&&
            $E_{p}(Z)$ & $E_{p}(N)$ & $D_{0}$  & $Q_{20}$ & $Q_{30}$ \cr
\cline{3-7} \cline{10-14} 
&& \multicolumn{6}{c}{D1S} &  \multicolumn{6}{c}{D1M}  \cr

$^{146}$Sm  && -14.64  & -3.58  & 0.00  &  1.20  &  0.00 &&& 
              -15.67  & -5.20  & 0.00  &  0.60  &  0.00  	 \\ 
$^{148}$Sm  && -12.41  & -3.34  & 0.18  &  3.00  &  1.25 &&& 
              -13.92  & -4.52  & 0.14  &  3.00  &  0.75          \\ 
$^{150}$Sm  && -10.98  & -1.53  & 0.41  &  4.80  &  1.50 &&& 
              -11.80  & -2.99  & 0.35  &  4.80  &  1.25          \\ 
$^{152}$Sm  &&  -6.58  & -5.57  & 0.00  &  7.20  &  0.00 &&&  
               -8.24  & -6.18  & 0.00  &  6.60  &  0.00          \\ 
$^{154}$Sm  &&  -5.91  & -3.27  & 0.00  &  7.80  &  0.00 &&& 
               -6.38  & -4.63  & 0.00  &  7.80  &  0.00          \\ 
\hline
\end{tabular}
\end{table}

From a theoretical perspective, many different models  have  been 
used to describe octupole correlations in atomic nuclei. For a 
detailed survey the reader is referred, for example, to 
Ref. \cite{butler_96}. Calculations based on the shell-correction 
approach with folded Yukawa deformed potentials \cite
{moller_81,leander_82}, as well as calculations based on Woods-Saxon 
potentials with various models for the microscopic and macroscopic 
terms \cite{naza_84,naza_92}, predicted a significant stabilization 
of octupole deformation effects  in various nuclear mass regions. 
Pioneer Skyrme-HF+BCS calculations including the octupole constraint
and restoring parity symmetry were carried out in Ref. \cite{mar83}.
Subsequent calculations in Ref. \cite{bon86} included both quadrupole 
and octupole constraints at the same time but at the mean field level only. On the 
other hand, microscopic studies of octupole correlations with Skyrme 
and Gogny EDFs, both at 
the mean field level and beyond  with different levels of 
complexity, have already been reported (see, Refs. \cite
{bon88,bon91,hee94,mey95,rob87,rob88,egi90,egi91,gar98,egi92,rob10} and references therein) 
for   several regions of the nuclear chart. Theoretical studies in 
the Sm region  include mean field based calculations with the collective 
hamiltonian and the Gogny force \cite{egi92},  the 
IBM study  with $spdf$ bosons of Ref. \cite{babilon_05} or the collective models using a 
coherent coupling between quadrupole and octupole modes \cite
{minkov_06} and new parametrizations of the quadrupole and octupole 
modes \cite{bizzeti_10}. Non-axial 
pear-like shapes in this region  were considered, for example, in 
Refs. \cite{skalski_91}. Additionally, the isotopes 
$^{146-156}$Sm have been investigated very recently within the 
constrained reflection-asymmetric relativistic mean field (RMF) 
approach \cite{zhang_10} based on the parametrization PK1 \cite
{RMF-PK1} for the RMF Lagrangian together with a constant gap BCS 
approximation for pairing correlations.

In the present work, we investigate the interplay between octupole 
and quadrupole degrees of freedom in the sample of nuclei 
$^{146-154}$Sm and $^{148-156}$Gd. We use three different levels of 
approximation. First, the constrained (reflection-asymmetric) 
Hartree-Fock-Bogoliubov (HFB) framework is used as starting point 
providing energy contour plots in terms of   the (axially symmetric) 
quadrupole $Q_{20}= \langle {\Phi} | \hat{Q}_{20} |  {\Phi} \rangle$ 
and octupole $Q_{30}= \langle {\Phi} | \hat{Q}_{30} |  {\Phi} 
\rangle$ moments (where $|  {\Phi} \rangle$ is the 
corresponding HFB intrinsic wave function). Within this mean field 
framework we  pay  attention to the shape changes in the considered 
nuclei and their relation with the underlying single-particle 
spectrum \cite{butler_96,rob10,ours-PT}.

As will be discussed  later on, the $(Q_{20},Q_{30})$ mean field 
potential energy surfaces (MFPES) obtained for the nuclei 
$^{146-154}$Sm and $^{148-156}$Gd are, in most of the cases, very 
soft along  the $Q_{30}$ direction indicating that the (static) mean 
field picture is not enough and that a (dynamical) beyond mean field 
treatment is required. Therefore, both the minimization of
the energies obtained after parity projection of the intrinsic states
\cite{mar83,egi91,egi92} as well as 
quadrupole-octupole configuration mixing calculations in the 
spirit of the Generator Coordinate Method (GCM) \cite{rs}, are 
subsequently carried out. The analysis of the two sets of results
allows to disentangle the role played in the dynamics of 
the considered nuclei by  the restoration of the broken 
reflection symmetry and the fluctuations in the ($Q_{20}$,$Q_{30}$) 
collective coordinates. Similar calculations with the Skyrme functional
where carried out in Ref \cite{mey95} for a lead isotope.

\begin{table} 
\label{tableGd}
\caption{The same as Table I but for  the isotopes $^{148-156}$Gd.
}
\begin{tabular}{cccccccccccccc}
\hline
Nucleus    && $E_{p}(Z)$ & $E_{p}(N)$ & $D_{0}$  & $Q_{20}$ & $Q_{30}$ &&&
              $E_{p}(Z)$ & $E_{p}(N)$ & $D_{0}$  & $Q_{20}$ & $Q_{30}$ \cr
\cline{3-7} \cline{10-14} 
&& \multicolumn{6}{c}{D1S} &  \multicolumn{6}{c}{D1M}  \cr

$^{148}$Gd  && -15.22  & -4.27  & 0.00  &  0.66  &  0.00 &&& 
              -16.11  & -5.35  & 0.00  &  0.00  &  0.00  	 \\ 
$^{150}$Gd  && -14.22  & -3.63  & 0.19  &  3.60  &  0.75 &&& 
              -15.43  & -5.03  & 0.05  &  3.00  &  0.25          \\ 
$^{152}$Gd  && -12.69  & -3.02  & 0.27  &  4.80  &  1.00 &&& 
              -13.18  & -4.76  & 0.15  &  4.80  &  0.50          \\ 
$^{154}$Gd  && -7.63   & -6.26  & 0.00  &  7.20  &  0.00 &&& 
              -9.25   & -6.88  & 0.00  &  6.60  &  0.00          \\ 
$^{156}$Gd  && -7.18   & -4.86  & 0.00  &  7.80  &  0.00 &&& 
              -7.66   & -6.20  & 0.00  &  7.80  &  0.00          \\ 
\hline
\end{tabular}
\end{table}

To the best of our knowledge, the hierarchy of approximations (i.e., 
reflection-asymmetric HFB, parity projection and $(Q_{20},Q_{30})$-GCM) 
considered in the present work belong, at least for the case 
of the Gogny-EDF, to the class of unique and state-of-the-art tools 
for the microscopic description of quadrupole-octupole correlations 
in atomic nuclei. Let us also stress that, the two-dimensional GCM 
(2D-GCM) framework used in the present study represents an extension 
of the treatment of octupolarity  reported in Refs. \cite
{rob10,egi92}, where a one-dimensional 
collective hamiltonian  based on several  approximations  and 
parameters extracted from $Q_{30}$-constrained HFB calculations, was 
considered. Here, on the other hand, the octupole and quadrupole 
degrees of freedom are explored simultaneously and the  kernels 
involved in the solution of the corresponding Hill-Wheeler equation 
\cite{rs} are  computed without assuming a gaussian behavior of the 
norm overlap neither a (second order)  expansion over the 
non-locality of the hamiltonian kernel. Therefore, the present study 
for the selected set of Sm and Gd nuclei, to the best of our 
knowledge the first of this kind for the case of the Gogny-EDF, may 
also be regarded as a proof of principle concerning the feasibility 
of the calculations to be discussed later on. Pioneer calculations 
along the same lines considered in the present study, but based on 
the Skyrme-EDF, have been carried out in Ref. \cite{mey95,Heenen.01}.

\begin{figure*}
\includegraphics[width=0.87\textwidth]{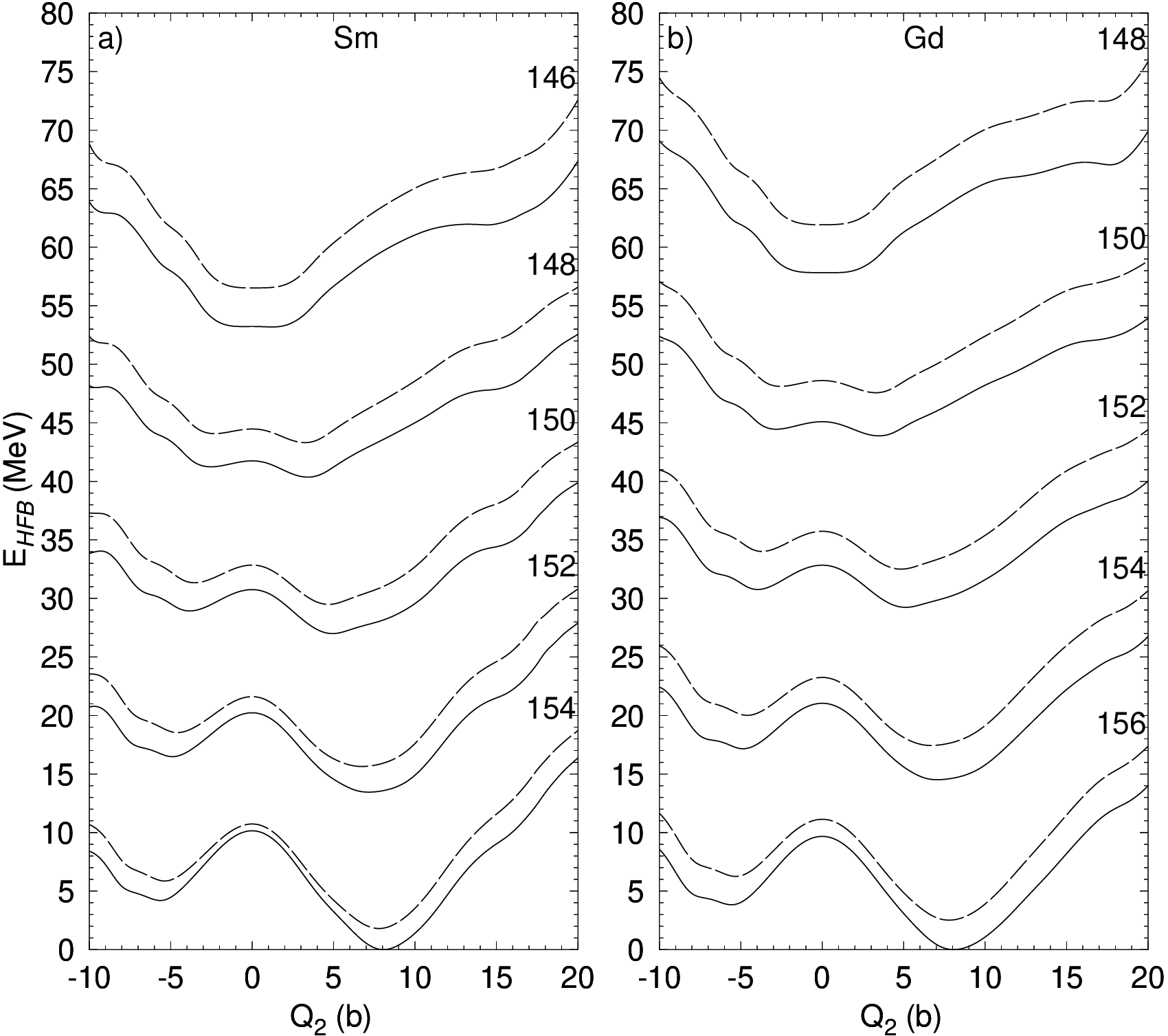} 
\caption{In panel a) the reflection symmetric (i.e., $Q_{30}$=0) MFPECs
 for $^{146-154}$Sm  and in panel b) for  
$^{148-156}$Gd are plotted as functions of the axially symmetric quadrupole moment 
$Q_{20}$. Results for both Gogny-D1S with full line and Gogny-D1M with dashed line are given.
In each panel the energies are referred to the D1S ground state energy of the heavier isotope.}
\label{figquadrupole}  
\end{figure*}

In addition to the standard Gogny-D1S \cite{gogny-d1s} 
parametrization, which is taken as a reference, the D1M 
parametrization \cite{gogny-d1m} will also be considered. The 
functional Gogny-D1S has a long standing tradition and it has been 
able to describe many low-energy experimental data all over the 
nuclear chart with reasonable predictive power both at the mean 
field level and beyond (see, for example, Refs. \cite{gogny-d1s, 
rob87,rob88,egi90,egi91,gar98,
rod02,egi04,egi04b, 
gogny-other-1,gogny-other-2,gogny-other-5,bertsch,peru,hilaire,delaroche10,
PLB-2010-rayner} and references therein). On the other hand, 
the D1M parametrization \cite{gogny-d1m} that was tailored to 
provide a better description of masses is now proving its merits in 
nuclear structure studies not only in even-even nuclei  \cite
{ours-PT,gogny-d1m,ours2,PLB-2010-rayner,ours-SrZrMo-quasi,ours-Rb-quasi,
ours-Y-Nb-quasi}, but also in odd nuclei in the framework of the 
Equal Filling Approximation (EFA) \cite 
{PLB-2010-rayner,ours-SrZrMo-quasi,ours-Rb-quasi,ours-Y-Nb-quasi}. 
In this paper the results of both D1S and D1M are compared to
verify the robustness of our predictions with respect to the 
particular version of the interaction and  to test the 
performance of D1M in the present context of quadrupole-octupole 
coupling.

The paper is organized as follows. In Secs. \ref{MF-RESULTS}, \ref
{BYMF-RESULTS-PP} and  \ref{BYMF-RESULTS-GCM} we  briefly describe  
the theoretical formalisms used in the present  work and 
subsequently the results obtained with them. Mean field 
calculations will be discussed in Sec. \ref{MF-RESULTS}. 
Parity projection and configuration mixing results will 
be presented in Secs. \ref{BYMF-RESULTS-PP}  and \ref
{BYMF-RESULTS-GCM}, respectively. In particular, in  Sec. \ref
{BYMF-RESULTS-GCM}  especial attention will be paid to beyond mean 
field properties in the considered nuclei -dynamical octupole and 
dipole moments, correlation energies, reduced transition 
probabilities B(E1) and B(E3) as well as energy splittings-  and 
their comparison with available experimental data. Finally, Sec. \ref
{CONCLU} is devoted to the concluding remarks and work perspectives.

\begin{figure*}
\includegraphics[width=0.91\textwidth]{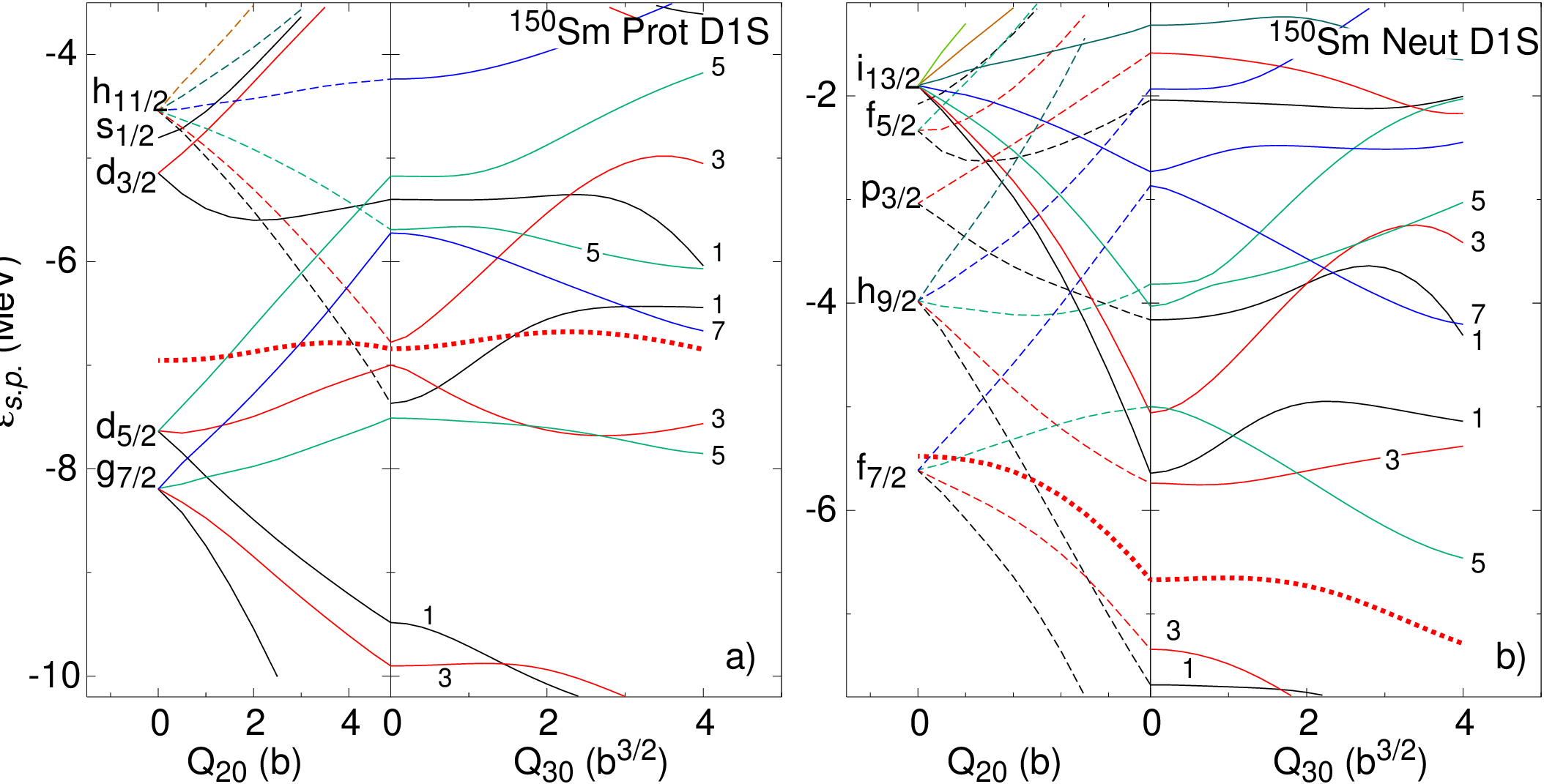}

\includegraphics[width=0.91\textwidth]{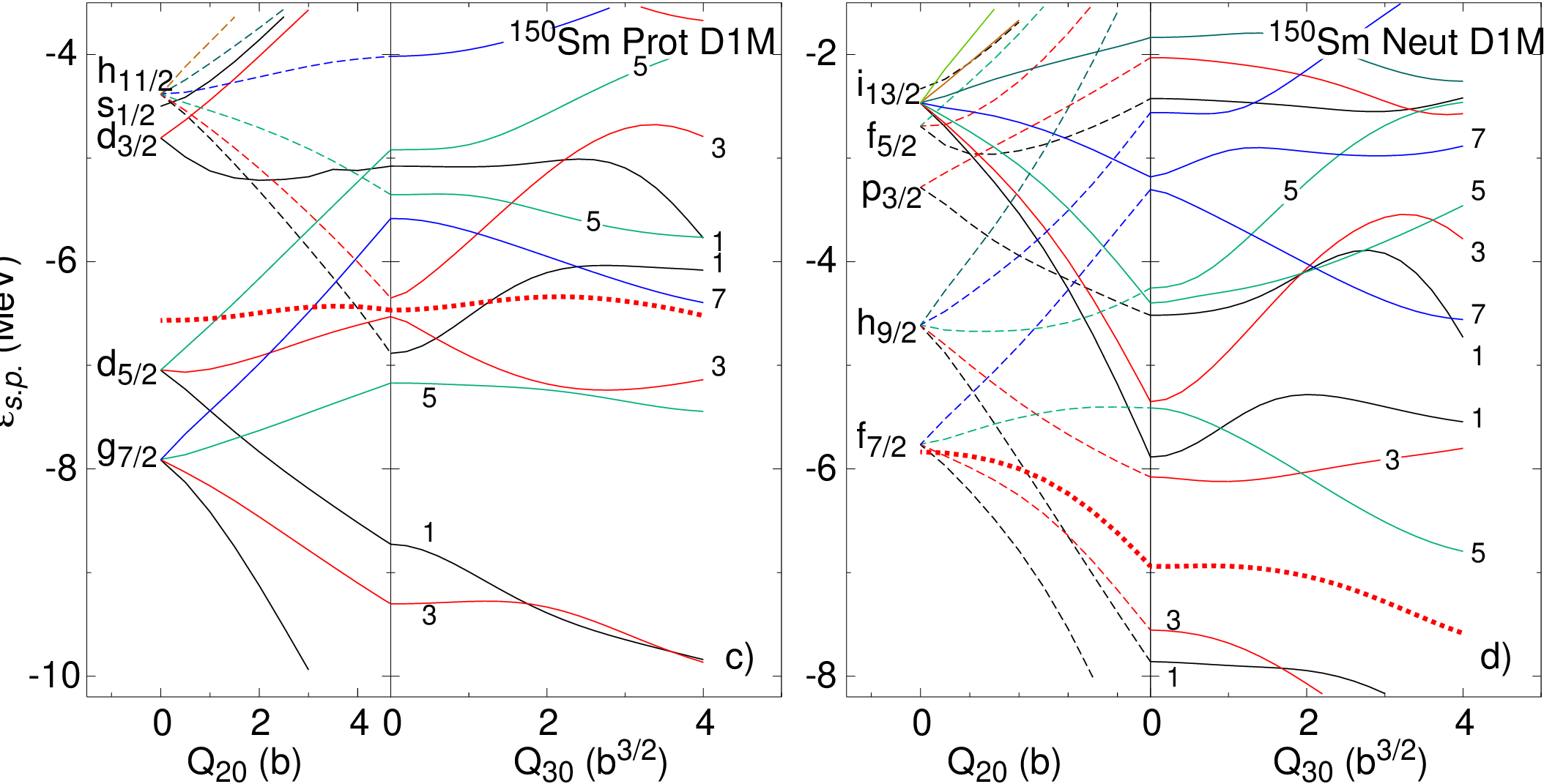}
\caption{(Color online) Single particle energies (see text for details)
in $^{150}$Sm
are plotted as a function of the quadrupole moment $Q_{20}$ (in barn) 
up to the value corresponding to the ground state minimum. From there on,
the plot continues with the representation of the SPEs as a function of the octupole moment
$Q_{30}$. In the part of the plot where the SPEs are plotted versus $Q_{20}$, 
full (dashed) curves stand for positive (negative) parity levels. 
The thick (red) dashed line in each plot represents the chemical
potential. In the part of the plot where the SPEs are plotted versus $Q_{30}$ 
some levels around the Fermi level are labeled with twice their $J_z$ value. 
Panels a) and b) (c) and d)) correspond to  results obtained with D1S (D1M) EDFs. 
Panels a) and c) (b) and d)) correspond to protons (neutrons). }
\label{fig_spe}
\end{figure*}



\section{Mean field systematics.}
\label{MF-RESULTS}

The aim of the present work is the study of the quadrupole-octupole 
dynamics in selected  Sm and Gd isotopes with neutron number 84 $\le$
N $\le$ 92. Three different levels of approximation are considered: 
the HFB method with constraints in the relevant degrees of freedom, 
parity projection (with minimization of the energy after projection) 
and the Generator Coordinate Method (GCM) with $Q_{20}$ and $Q_{30}$ as 
collective coordinates. For a detailed survey on the three 
techniques the reader is referred to Ref. \cite{rs}. The Gogny EDF 
is used consistently in the three methods both with the D1S and D1M 
parametrizations.

First, $(Q_{20},Q_{30})$-constrained HFB calculations are performed 
for the nuclei $^{146-154}$Sm and $^{148-156}$Gd to 
obtain a set of states $| \Phi ({\bf Q}) \rangle$ labeled by their 
corresponding multipole moments ${\bf Q}=(Q_{20},Q_{30})$. The $K=0$ 
quadrupole $Q_{20}$ and octupole  $Q_{30}$ moments 
are given by the average values 
\begin{equation} Q_{20} = \langle 
\Phi | z^{2} - \frac{1}{2}\left(x^{2} + y^{2} \right) | \Phi \rangle,
\end{equation} 
and 
\begin{equation} Q_{30} = \langle \Phi | z^{3} - 
\frac{3}{2}\left(x^{2} + y^{2} \right)z | \Phi \rangle.
\end{equation}
Axial and  time reversal are self-consistent symmetries in the mean 
field  calculations. As a consequence of the axial symmetry imposed 
on the HFB wave functions $|  {\Phi}\rangle$, the mean values of the 
multipole operators $\hat{Q}_{2 \mu}$ and $\hat{Q}_{3 \mu}$ with 
$\mu \ne 0$ are zero by construction. Aside from the constraints on 
the quadrupole and octupole moments, a constraint on the center of 
mass operator is used to place it at the origin of coordinates in order 
to prevent spurious effects associated to center of mass motion. The HFB 
quasiparticle operators $(\hat{\alpha}_{k}^{\dagger}, 
\hat{\alpha}_{k})$ \cite{rs} have been expanded in an axially 
symmetric harmonic oscillator (HO) basis $(\hat{c}_{l}^{\dagger}, 
\hat{c}_{l})$ containing 13 major shells as to grant 
convergence for all the observable quantities. For the solution of 
the HFB equation, an approximate second order gradient method \cite
{rob11} is used. 

The MFPES have been computed 
in a grid with $Q_{20}$ in the range from -30b to 30b in steps of 
0.6 b and the octupole moment $Q_{30}$ in the range from 0 b$^{3/2}$ to 3.75 b
$^{3/2}$ in steps of 0.25 b$^{3/2}$. Negative values of the octupole 
moment are not computed explicitly as the corresponding wave 
function can be obtained from the positive $Q_{30}$ one by 
applying the parity operator. As the Gogny EDF is invariant under 
parity (see \cite{rod02,egi04} for a discussion 
of the meaning of symmetry invariance for density dependent 
"forces") the energy has the property 
$E_{HFB}(Q_{20},Q_{30})=E_{HFB}(Q_{20},-Q_{30})$ and therefore is an 
even function of the octupole moment. For this reason, in the 
graphical representation of the PES only positive values of $Q_{30}$ 
are considered.

\begin{figure*}
\includegraphics[width=0.508\textwidth]{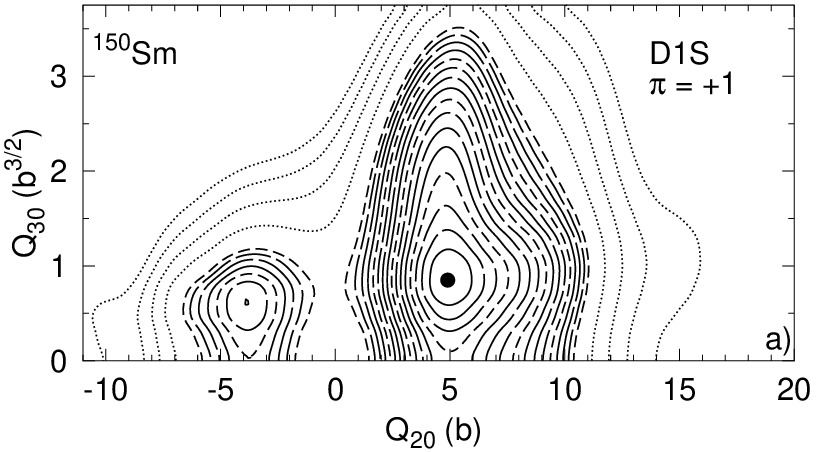}
\includegraphics[width=0.480\textwidth]{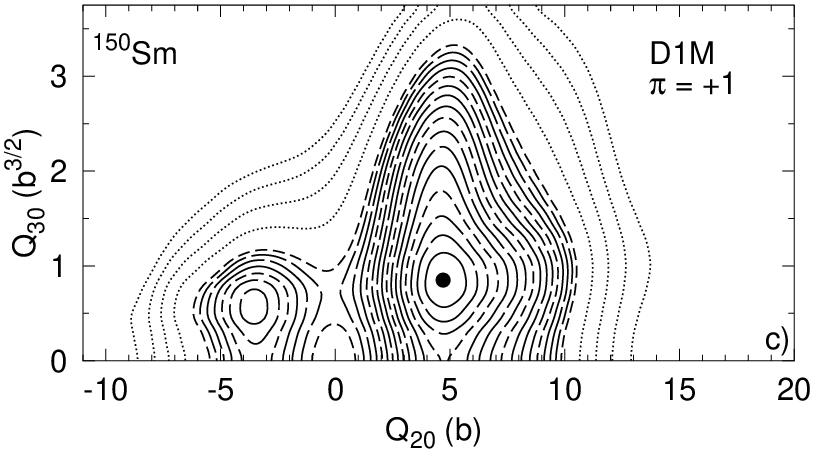} \par
\includegraphics[width=0.508\textwidth]{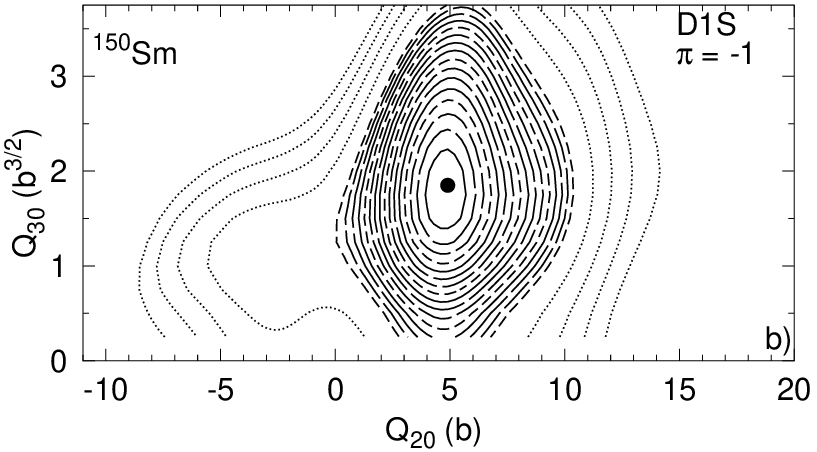}
\includegraphics[width=0.480\textwidth]{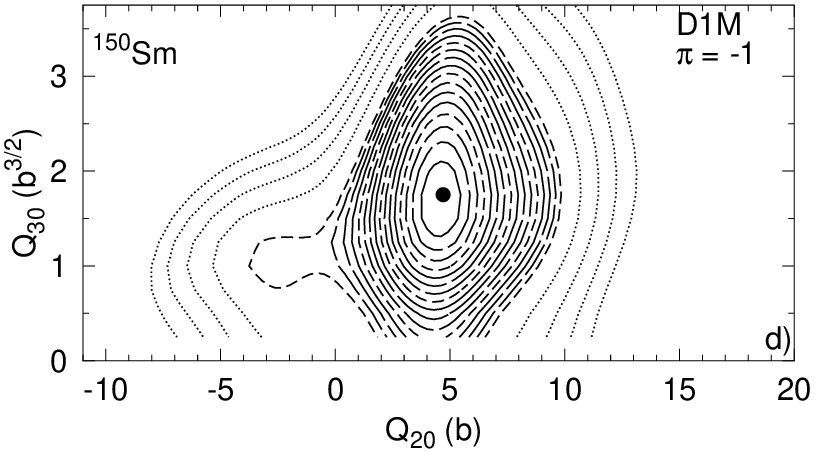} \par
\caption{Positive $\pi$=+1 (upper panels) and negative $\pi$=-1
(lower panels)
parity-projected potential energy surfaces (PPPES) computed
with the Gogny-D1S (left panel) and Gogny-D1M
(right panel) EDFs for the nucleus $^{150}$Sm. See caption of Fig \ref{fig_mfq2q3_Sm} for
the contour line patterns.}
\label{fig_ppq2q3_Sm}
\end{figure*}

The MFPESs obtained for the  nucleus $^{150}$Sm, with the 
parametrizations D1S and D1M of the Gogny-EDF, are shown in Fig. \ref
{fig_mfq2q3_Sm} as an illustrative example of our mean field 
results. For the sake of presentation, quadrupole and octupole 
moments have  been constrained in the plots to the ranges -10 b 
$\le Q_{20} \le$ 20 b and 0 b$^{3/2}$ $\le Q_{30} \le$ 3.75 b$^{3/2}$,
respectively. The similitude between the D1S 
and D1M results both in the $Q_{20}$ and $Q_{30}$ directions is remarkable. In 
previous calculations in other regions and looking at different 
physical effects \cite
{ours-PT,PLB-2010-rayner,ours-SrZrMo-quasi,ours-Rb-quasi,
ours-Y-Nb-quasi} we have already noticed the same similitude between D1S and 
D1M results. Focusing on the MFPES, the absolute minimum is located 
in the prolate side at a finite value of the octupole moment. The 
minimum is very shallow along the $Q_{30}$ direction. Another minimum is observed in the oblate side, but 
this time centered at $Q_{30}=0$. For the other nuclei considered 
the energies look similar and therefore they are not shown. The most relevant mean field 
quantities for the ground states are summarized  in Tables  I and II. 
In order to better understand the quadrupole deformation properties of the 
studied nuclei, the reflection symmetric (i.e., $Q_{30}$=0) mean 
field potential energy curves (MFPECs) are depicted for all the 
considered nuclei in Fig. \ref{figquadrupole}. A transition 
from weakly deformed ground states in the N=84 nuclei $^{146}$Sm and 
$^{148}$Gd  to well (quadrupole) deformed ground states in 
$^{152,154}$Sm and $^{154,156}$Gd (prolate moments 6.6 b $\le 
Q_{20} \le$ 7.8 b) is observed. In most of the isotopes except the 
lightest ones an additional minimum is observed in the oblate side. 
This minimum  may become a saddle point (see, \cite {ours_x5_2} for 
examples) once the $\gamma$ degree of freedom is considered. 
Nevertheless, the simultaneous consideration of triaxial quadupole 
and octupole moments lies outside of the scope of the present 
study. Investigation along these lines is in progress and will be 
reported elsewhere.

\begin{figure*}
\includegraphics[width=0.5\textwidth]{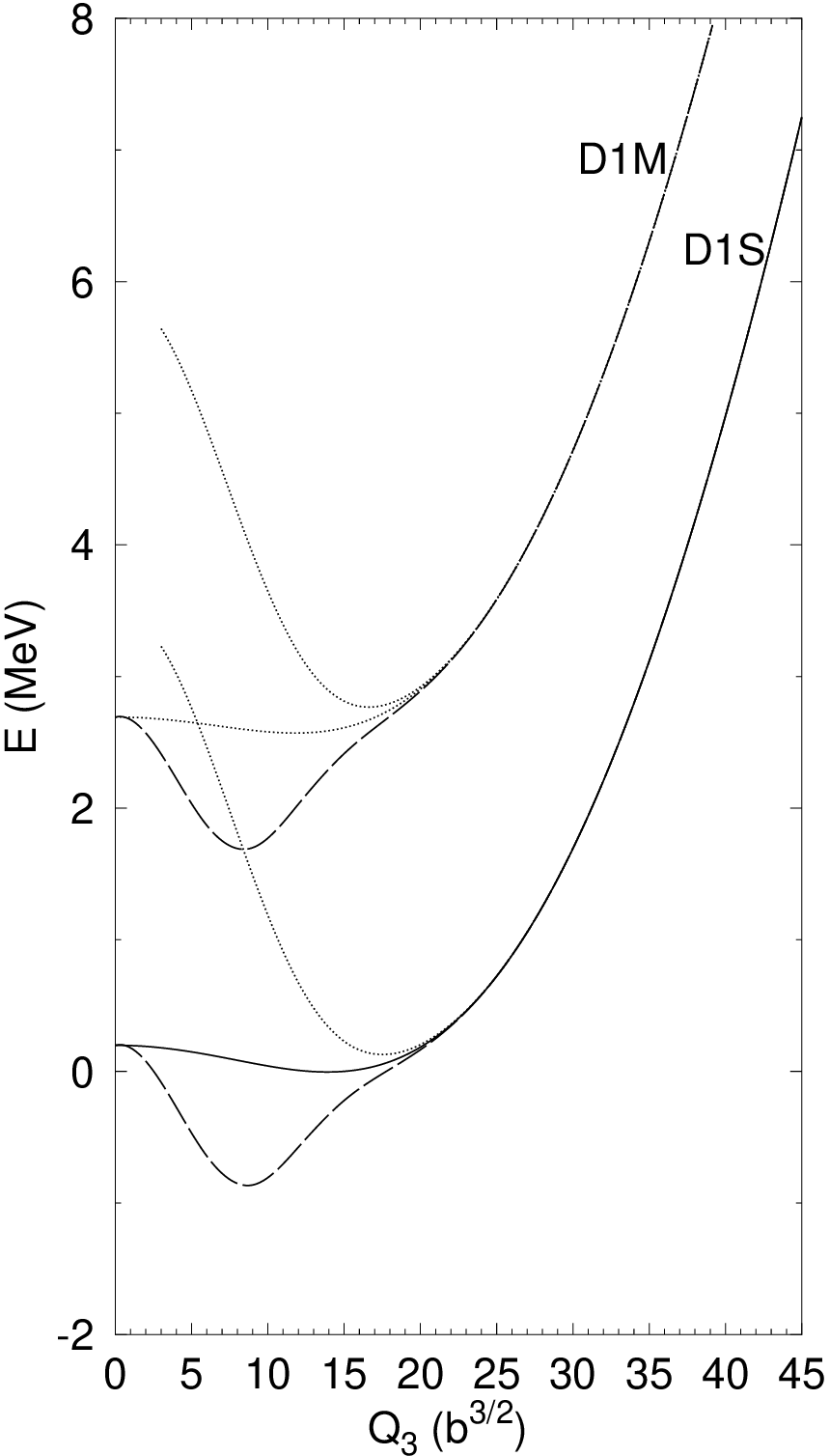}
\caption{Mean field (full line), positive  (dashed)
and negative (dotted) parity-projected energies as a function of
the octupole moment $Q_{30}$ for selfconsistent $Q_{20}$ values 
for the nucleus $^{150}$Sm. Energies are referred to the D1S HFB ground
state energy.}
\label{evol-min-gamma-Sm}
\end{figure*}

From  Tables  I and II, we observe  the onset of an 
octupole deformed regime at the N=88 nuclei $^{150}$Sm and 
$^{152}$Gd. These nuclei mark the borders of  another shape
transition from  octupole deformed ground states in  $^{148}$Sm and 
$^{150}$Gd to quadrupole deformed and reflection symmetric ground 
states in $^{152}$Sm and $^{154}$Gd. Consistent with the breakdown 
of the left-right symmetry in their ground states, the $^{148,150}$Sm 
and $^{150,152}$Gd isotopes exhibit a non zero  (static) dipole 
moment $D_0$. It is computed as the ground state average value of the 
dipole operator 
\begin{equation} \label{dipole-operator}
\hat{D}_{0} = \frac{N}{A} \hat{z}_{prot} - \frac{Z}{A} \hat{z}_{neut} 
\end{equation}
along the symmetry z-axis. The values of $D_0$ tend to be smaller 
for D1M than for D1S. This is not surprising due to the delicate 
balance between single particle orbital properties that enter in the 
definition of the dipole moment \cite{egi90}. Another quantity of 
interest is the mean field octupole correlation energy 
$ E_{corr}^{MF} = E_{HFB,Q_{30}=0}^{g.s} - E_{HFB}^{g.s} $ corresponding
to the energy gain by allowing octupole deformation. 
For example, the values obtained for  $^{150}$Sm and $^{152}$Gd are 
204 and 43 KeV (105 and 6 KeV) for the  functional D1S (D1M), 
respectively. These very low values are a clear indication of the 
softness of the octupole minima in those nuclei. As the minima are 
also soft along the $Q_{20}$ direction both the quadrupole and 
octupole degrees of freedom have to be considered at the same time 
in a dynamical treatment of the problem \cite{urban_87,urban_91,garrett_09}.

\begin{figure*}

\includegraphics[width=0.45\textwidth]{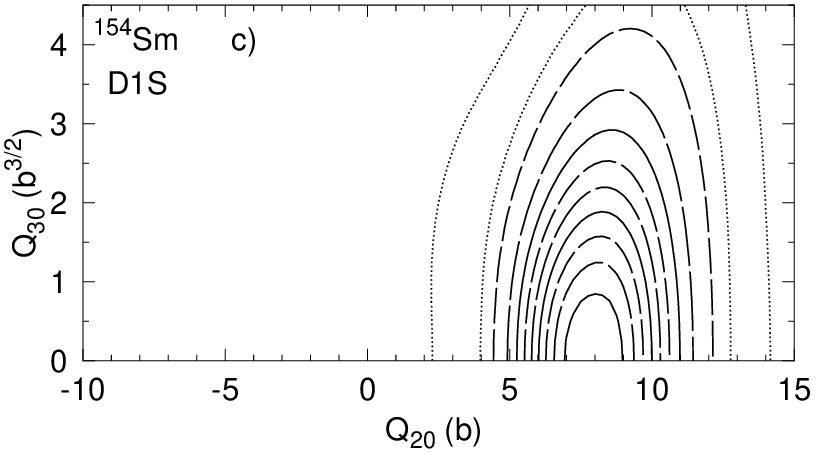}
\includegraphics[width=0.45\textwidth]{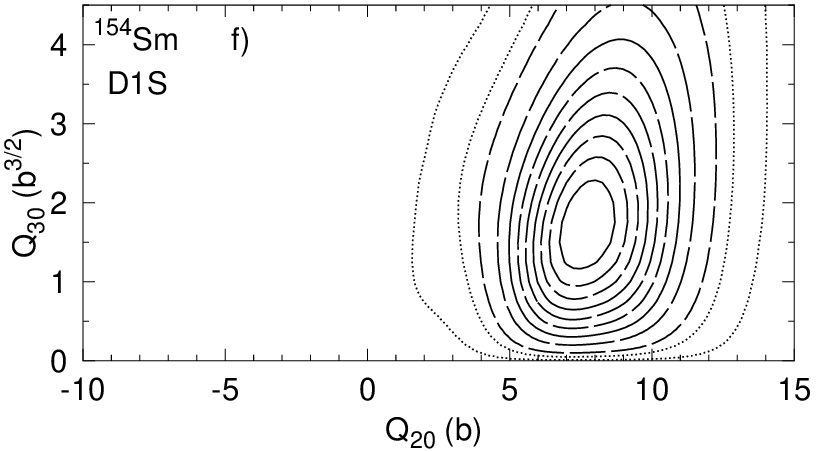} \par
\includegraphics[width=0.45\textwidth]{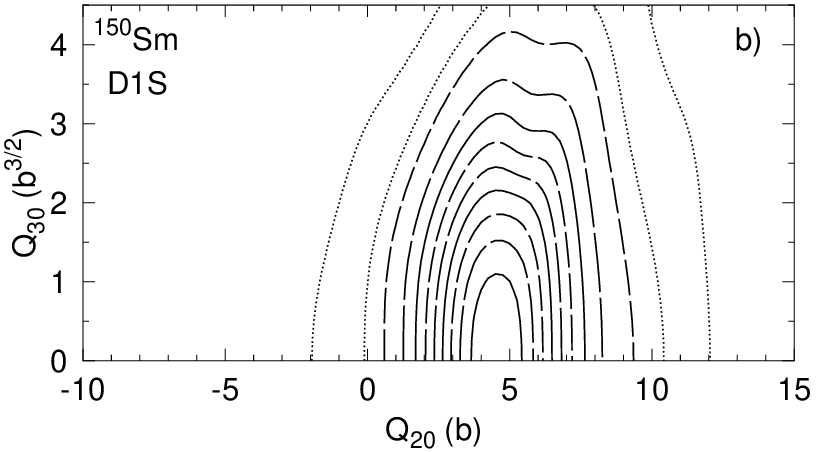}
\includegraphics[width=0.45\textwidth]{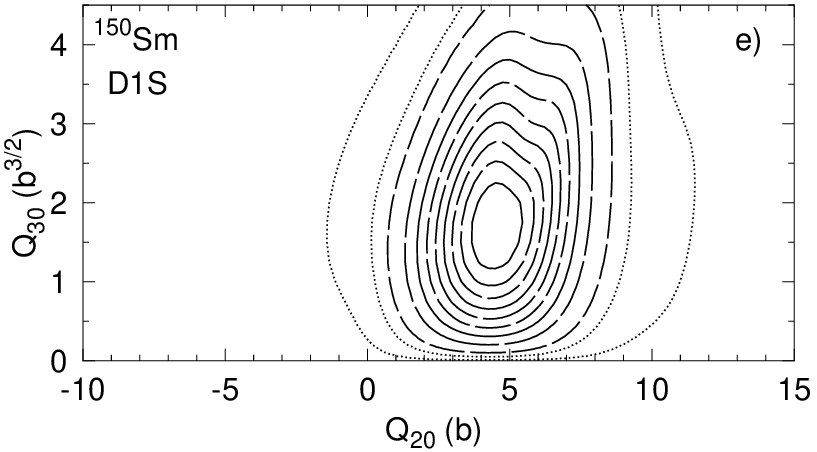} \par
\includegraphics[width=0.45\textwidth]{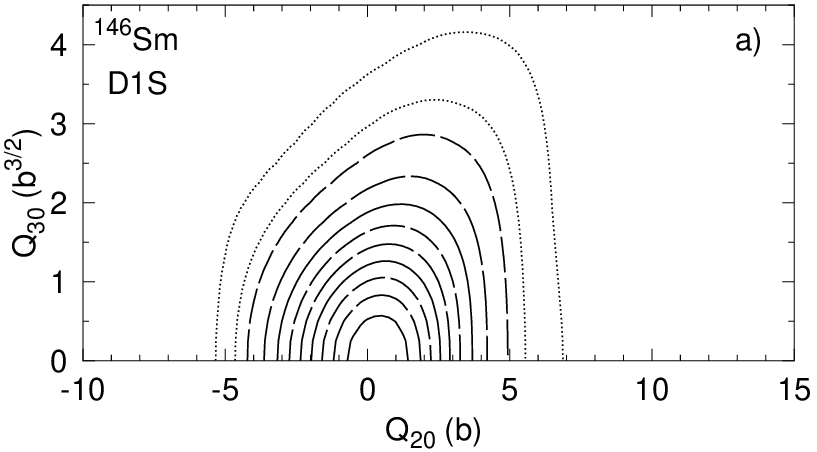}
\includegraphics[width=0.45\textwidth]{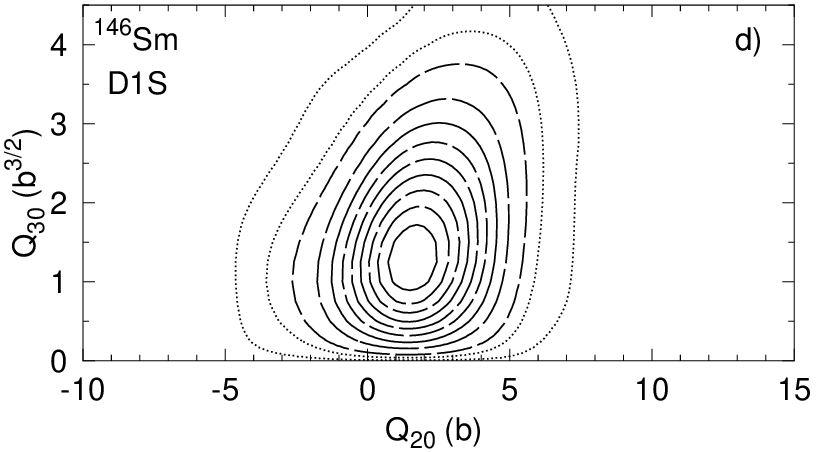} \par%
\caption{Collective wave functions squared  
($| G_{\sigma} (Q_{20},Q_{30})|^{2}$) for the ground state (left panels) and
the lowest negative parity state (right panels) for the nuclei
 $^{146,150,154}$Sm. The contour lines (a sucession of full, long dashed and
short dashed) start at 90 \% of the maximum value up to 10 \% of it. The two
dotted line contour correspond to the tail of the amplitude (5 \% and 1\% of the maximum
value).
}
\label{coll-wfs-GCMQ2Q3-Sm}
\end{figure*}

In Tables I and  II,  the proton $E_{p}(Z)$ and neutron $E_{p}(N)$ 
pairing energies are also listed. They are computed 
in the usual way as $E_{p} (\tau) =-1/2 \textrm{Tr} \Big(\Delta (\tau) \kappa^{*} (\tau) \Big)$ 
in terms of the pairing field $\Delta$ and the pairing tensor $\kappa$ 
for each isospin $\tau$ = Z, N. Moving along isotopic chains, 
the smallest neutron pairing energy corresponds to the N=88 
nuclei $^{150}$Sm and $^{152}$Gd, which are precisely the ones 
providing the largest values of the mean field octupole correlation 
energy $E_{corr}^{MF}$. The significant lowering of the neutron 
pairing energies in these nuclei is a  consequence of the low 
level density typical of 
deformed (quadrupole or octupole) minima, the Jahn-Teller effect. On the other hand, 
proton pairing energies tend to decrease as a function of the 
neutron number. In general, the proton and neutron pairing energies 
for the two Gogny-EDFs considered follow the same trend, the only 
relevant difference being in their  absolute values that tend to be 
slightly larger for D1M.

\begin{table} 
\label{tableSm_Q2_Q3_GCM}
\caption{Dynamical quadrupole $\bar{Q}_{20}^{(+)}$, $\bar{Q}_{20}^{(-)}$ (b)
and octupole $\bar{Q}_{30}^{(+)}$, $\bar{Q}_{30}^{(-)}$
(b$^{3/2}$  ) moments corresponding to the first  positive and negative  parity
2D-GCM  states in the isotopes $^{146-154}$Sm. Results are given for both Gogny-D1S and 
Gogny-D1M EDFs.
}
\begin{tabular}{cccccccccccc}
\hline
Nucleus && $\bar{Q}_{20}^{(+)}$  & $\bar{Q}_{30}^{(+)}$ & 
           $\bar{Q}_{20}^{(-)}$  & $\bar{Q}_{30}^{(-)}$ & & &
           $\bar{Q}_{20}^{(+)}$  & $\bar{Q}_{30}^{(+)}$ & 
           $\bar{Q}_{20}^{(-)}$  & $\bar{Q}_{30}^{(-)}$ \\ 
\cline{3-6} \cline{9-12} 
&& \multicolumn{4}{c}{D1S} &&&  \multicolumn{4}{c}{D1M}  \cr

$^{146}$Sm && 0.63 & 0.43  &  1.37  & 1.33 &&&
             0.45 & 0.39  &  1.39  & 1.29   	 \\  
$^{148}$Sm && 2.28 & 0.54  &  2.01  & 1.77 &&&
             2.93 & 0.52  &  2.10  & 1.65        \\  
$^{150}$Sm && 2.94 & 0.60  &  2.63  & 1.83 &&&
             3.09 & 0.56  &  2.68  & 1.81      	 \\ 
$^{152}$Sm && 5.48 & 0.51  &  3.81  & 1.72 &&&
             5.28 & 0.50  &  3.63  & 1.74        \\
$^{154}$Sm && 6.15 & 0.50  &  4.21  & 1.63 &&&
             6.15 & 0.49  &  4.33  & 1.58        \\   
\hline
\end{tabular}
\end{table}

Before concluding this section, we turn our attention to 
single-particle properties. The appearance of  quadrupole and/or 
octupole deformation effects is strongly linked to the position of 
the Fermi energy in the single-particle spectrum \cite 
{butler_96,rob10,ours2,ours-PT,Bo-M}. Therefore, the evolution 
of the the single-particle energies (SPEs) for both protons and 
neutrons with deformation is an interesting piece of information. In 
HFB calculations the concept of single particle energy is assigned 
to  the eigenvalues of the Routhian $h = t+ \Gamma - 
\lambda_{Q_{20}}Q_{20} - \lambda_{Q_{30}}Q_{30}$, with $t$ being the 
kinetic energy operator, and $\Gamma$ the Hartree-Fock field. The 
term  $\lambda_{Q_{20}}Q_{20} + \lambda_{Q_{30}}Q_{30}$ contains the 
Lagrange multipliers used to enforce the corresponding quadrupole 
and octupole constraints.

Proton and neutron  SPEs for the nucleus $^{150}$Sm, computed with 
both the Gogny-D1S and Gogny-D1M EDFs  are presented 
in Fig. \ref{fig_spe}.  The SPEs are plotted first
as  functions of the quadrupole moment $Q_{20}$ up to the value 
corresponding to the ground state minimum obtained with the 
$Q_{30}=0$ constraint. From there on, the plot continues with the 
representation of the SPEs as a function of the octupole moment 
$Q_{30}$.  The given SPEs as a function of the octupole 
moment have the self-consistently determined quadrupole moment
which, in the present case, does not depart 
significantly from the ground state value at $Q_{30}=0$.

\begin{table} 
\label{tableGd_Q2_Q3_GCM}
\caption{The same as Table III but for the isotopes 
$^{148-156}$Gd. 
}
\begin{tabular}{cccccccccccc}
\hline
Nucleus && $\bar{Q}_{20}^{(+)}$  & $\bar{Q}_{30}^{(+)}$ & 
           $\bar{Q}_{20}^{(-)}$  & $\bar{Q}_{30}^{(-)}$ & & &
           $\bar{Q}_{20}^{(+)}$  & $\bar{Q}_{30}^{(+)}$ & 
           $\bar{Q}_{20}^{(-)}$  & $\bar{Q}_{30}^{(-)}$ \\ 
\cline{3-6} \cline{9-12} 
&& \multicolumn{4}{c}{D1S} &&&  \multicolumn{4}{c}{D1M}  \cr  
$^{148}$Gd && 0.23 & 0.44  &  1.05  & 1.35 &&& 
             0.12 & 0.41  &  0.97  & 1.29        \\   
$^{150}$Gd && 2.46 & 0.52  &  1.78  & 1.74 &&& 
             2.53 & 0.47  &  1.57  & 1.65        \\   
$^{152}$Gd && 3.47 & 0.57  &  2.66  & 1.79 &&& 
             3.50 & 0.55  &  2.73  & 1.73        \\ 
$^{154}$Gd && 5.72 & 0.50  &  3.75  & 1.74 &&& 
             5.50 & 0.49  &  3.31  & 1.68        \\
$^{156}$Gd && 6.51 & 0.49  &  4.47  & 1.59 &&& 
             6.40 & 0.48  &  4.56  & 1.61        \\   
\hline
\end{tabular}
\end{table}

The first significant conclusion drawn from Fig. \ref{fig_spe} is 
that the D1S and D1M SPE plots look rather similar near the Fermi 
level (thick red dashed line): both the ordering of the levels at 
sphericity and their behavior with $Q_{20}$ and $Q_{30}$ are 
rather similar. For this reason we will from now on focus only on 
the D1M SPEs. For protons, the positive parity $d_{5/2}$ orbital 
strongly interacts with the negative parity $h_{11/2}$ one by means 
of the $l=3$ octupole component of the interaction. The position of the 
proton's Fermi level in the considered nucleus is located in the 
center of a small gap in the single particle spectrum that favors 
octupole deformation (Jahn-Teller effect \cite{JTE-1}). In the 
neutron's spectrum a fairly large gap near the Fermi level also 
opens up when the octupole moment is switched on. The neighboring 
levels come from the negative parity $f_{7/2}$ orbital and the 
positive parity $i_{13/2}$ intruder orbital. It is also worth 
mentioning the occurrence of  "quasi-$j$" orbitals in the neutron 
spectrum for the $Q_{30}$ values corresponding to the minimum at 
around 2 b$^{3/2}$. A $j=7/2$ is formed at an energy of around $-4$ 
MeV; one with $j=5/2$ is located at around $-6$ MeV and finally 
another one with $j=3/2$ shows up at an energy of $-8$ MeV. The same 
grouping of levels can also be observed in the SPEs for protons at 
similar values of the octupole moment. These quasi-$j$ orbitals are 
the consequence of the relationship between classical closed 
periodic orbits for specific octupole deformed shapes and the 
corresponding quantum orbitals that have to show an integer ratio 
between the radial and angular frequencies (see \cite{Bo-M}, Vol II, 
page  587 for a general discussion and also \cite{egi90} for 
specific examples in rare earth nuclei).

%
%

\section{Parity Projection.}
\label{BYMF-RESULTS-PP}

Although the HFB framework discussed in the previous section is a 
valuable starting point, it produces MFPESs with very soft minima along 
the $Q_{30}$-direction in the nuclei considered. This 
suggests the important role played by both types of dynamical correlations: 
the one associated with symmetry 
restoration and the other to configuration mixing.  
Symmetry restoration is considered in this section while 
configuration mixing will be presented in the  next section.

\begin{figure*}
\includegraphics[width=0.98\textwidth]{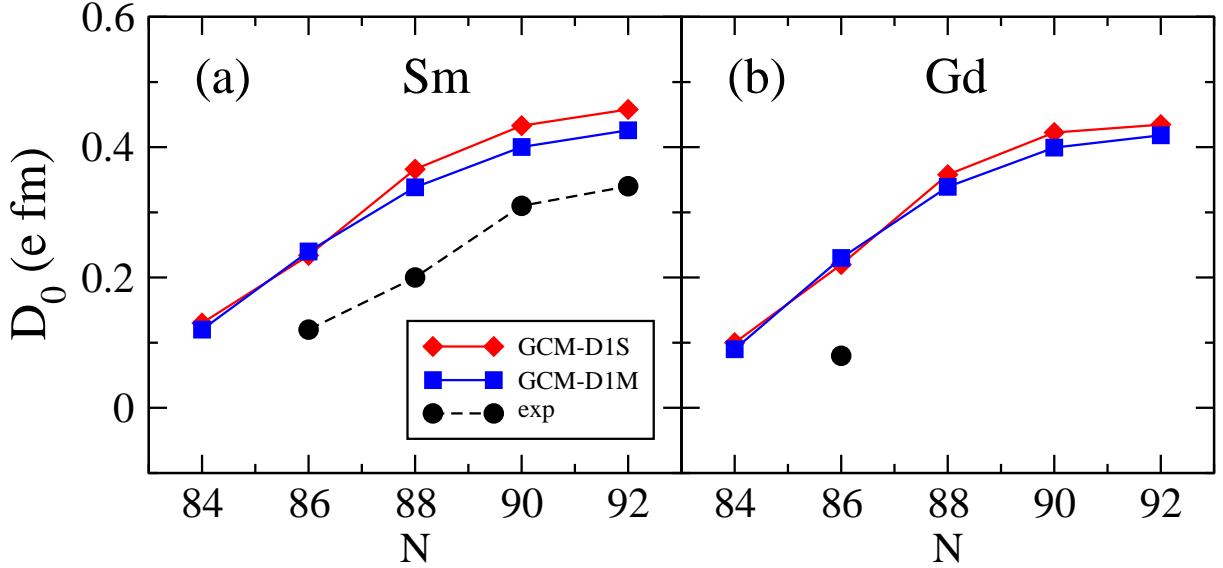}
\caption{(Color online) The dynamical 
dipole moments provided by the GCM calculations 
for the nuclei $^{146-154}$Sm and $^{148-156}$Gd
are shown, as functions of the neutron number, in panels a-b). Experimental dipole
moments are taken from Ref. \cite{butler_96}. Results are shown for both Gogny-D1S and Gogny-D1M
EDFs.}
\label{Dipole_GCM}
\end{figure*}

There are two spatial symmetries broken in the present calculations. 
One is rotational symmetry with the quadrupole moment as relevant 
parameter and the other is reflection symmetry (parity) with the 
octupole moment as relevant quantity. From the discussion of the 
mean field results it is clear that the softest mode is the octupole 
moment and therefore the most relevant symmetry to be restored is 
parity.
Obviously, it would be desirable to restore also the rotational 
symmetry as well as particle number. This combined symmetry 
restoration is feasible but, when combined with the configuration 
mixing of next section, becomes a very demanding computational task 
not considered in this paper.

The quantum interference typical of the GCM framework could be directly used to 
restore the parity  symmetry by choosing appropriate weights
for the configurations with mutipole moments $(Q_{20},Q_{30})$ and 
$(Q_{20},-Q_{30})$  \cite {egi92}. 
However, in order to disentangle the relative contribution of the 
parity restoration correlations  as compared with 
the ones of the  GCM configuration mixing, we have carried out  
explicit parity projection calculations.

To restore parity symmetry \cite{mar83,egi91} we build positive ($\pi = +1$) and negative  (
$\pi = -1$) parity-projected states $| {\Phi}^{\pi} (Q_{20},Q_{30}) 
\rangle = \hat{{\cal{P}}}^{\pi}| {\Phi} (Q_{20},Q_{30}) \rangle$ by applying
the parity projector $\hat{{\cal{P}}}^{\pi}$ to the intrinsic configuration.
The parity projector is a linear combination of the identity and the parity
operator $\hat{\Pi}$ given by
\begin{equation}
\hat{{\cal{P}}}^{\pi} = \frac{1}{2} \left(1 + \pi \hat{\Pi} \right).
\end{equation}
The projected energies, used to construct  
parity-projected potential energy surfaces (to be called PPPES in what follows), are
labeled with the multipole moments ${\bf Q}=(Q_{20},Q_{30})$ of the intrinsic state and
read \cite{rob07}
\begin{equation} \label{PROJEDF}
E_{\pi} ({\bf Q}) =
\frac{
\langle {\Phi} ({\bf Q}) | \hat{H} [\rho(\vec{r})] | {\Phi} ({\bf Q}) \rangle
}
{
\langle {\Phi} ({\bf Q}) |  {\Phi} ({\bf Q}) \rangle
+
\pi \langle {\Phi} ({\bf Q}) |  \hat{\Pi} | {\Phi} ({\bf Q}) \rangle
}
+ \pi
\frac{\langle {\Phi} ({\bf Q}) |
\hat{H} [\theta(\vec{r})]  \hat{\Pi} | {\Phi} ({\bf Q}) \rangle
}
{
\langle {\Phi} ({\bf Q}) |  {\Phi} ({\bf Q}) \rangle
+
\pi \langle {\Phi} ({\bf Q}) |  \hat{\Pi} | {\Phi} ({\bf Q}) \rangle
}
\end{equation}

The parity projected mean value of proton and neutron number, 
$\frac{\langle {\Phi} ({\bf Q}) |  \hat{Z} \hat{{\cal{P}}}^{\pi} | {\Phi} ({\bf Q}) \rangle
}{\langle {\Phi} ({\bf Q}) |\hat{{\cal{P}}}^{\pi}  |  {\Phi} ({\bf Q}) \rangle}$ 
and 
$\frac{\langle {\Phi} ({\bf Q}) | 
\hat{N} \hat{{\cal{P}}}^{\pi} | {\Phi} ({\bf Q}) \rangle}
{\langle {\Phi} ({\bf Q}) |\hat{{\cal{P}}}^{\pi}  |  {\Phi} ({\bf Q}) \rangle}$ 
usually differ from the nucleus' proton  $Z_{0}$ and neutron $N_{0}$ numbers. To correct the energy
for this deviation we have 
replaced   $\hat{H}$ by 
$\hat{H} -  \lambda_{Z} \left( \hat{Z} -Z_{0} \right)-  \lambda_{N} \left( \hat{N} -N_{0} \right)$, where 
$\lambda_{Z}$ and $\lambda_{N}$ are chemical potentials 
for protons and neutrons, respectively \cite{har82,bon90,egi91}.

\begin{figure*}
\includegraphics[width=0.98\textwidth]{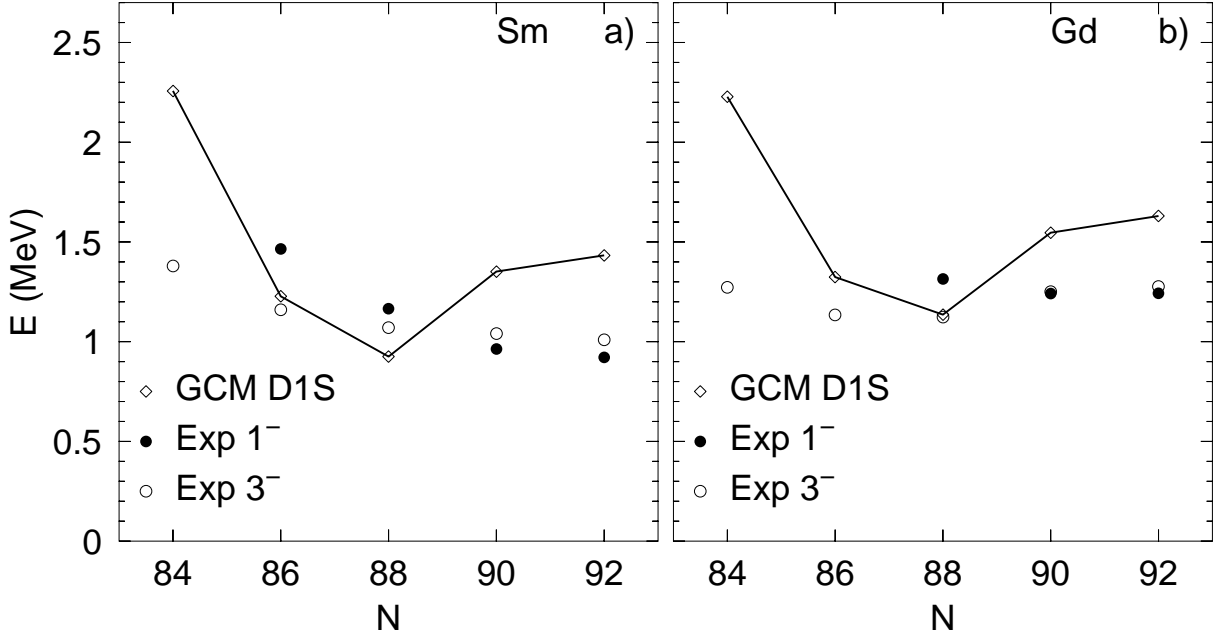}
\caption{Predicted energy splittings between the lowest lying $\pi$=+1 and 
$\pi$=-1  2D-GCM  states  in $^{146-154}$Sm and $^{148-156}$Gd are 
compared with  the experimental $0^{+}-1^{-}$ and $0^{+}-3^{-}$ 
splittings \cite{exp_ensdf}. Results are shown for the D1S parametrization
of the Gogny force as D1M ones are rather similar.
}
\label{splitting01_GCM}
\end{figure*}

In the case of the Gogny-EDF, as well as for Skyrme-like EDFs, the 
definite expression for the projected energy (\ref{PROJEDF}) 
depends, on the prescription used for the density dependent part of 
the functional. In this work, we resort to the so called {\it{mixed 
density}} prescription that amounts to consider the standard 
intrinsic density 
\begin{equation}
\rho(\vec{r})= 
\frac{
\langle {\Phi} ({\bf Q}) | \hat{\rho}({\vec{r})} | {\Phi} ({\bf Q}) \rangle
}
{
\langle {\Phi} ({\bf Q}) | {\Phi} ({\bf Q}) \rangle
},
\end{equation}
and the density 
\begin{equation}
\theta(\vec{r})= 
\frac{
\langle {\Phi} ({\bf Q}) | \hat{\rho}({\vec{r})} \hat{\Pi} | {\Phi} ({\bf Q}) \rangle
}
{
\langle {\Phi} ({\bf Q}) | \hat{\Pi} | {\Phi} ({\bf Q}) \rangle
}
\end{equation}
in the evaluation of the first and second terms in  Eq. (\ref{PROJEDF}), 
respectively. The {\it{mixed density}} prescription has been 
widely and successfully used in the context of projection  and/or 
configuration mixing  techniques (see, for example, \cite {bon90,rod02,egi04,review-Bender,delta2p-2,Duguet-1,Doba-mixed} 
and references therein). In fact, this is the 
only prescription that guarantees various consistency requirements 
within the EDF framework \cite{rod02,rob07,Duguet-kernels-1}. Even though 
this prescription has some drawbacks, as put into evidence recently \cite 
{Duguet-kernels-1,Duguet-kernels-2,Duguet-kernels-3}, the use of 
other prescriptions, like the one based on the projected density, 
are pathologically ill defined when applied to the restoration of spatial 
symmetries \cite{rob10b}.

As an illustrative example of PPPES, we show in Fig. \ref{fig_ppq2q3_Sm} the 
results for the nucleus $^{150}$Sm obtained with both the D1S and D1M 
parametrizations of the Gogny force. Along the $Q_{30}$=0 axis, the 
projection onto positive parity $\pi$=+1 is unnecessary since  the 
corresponding (quadrupole deformed) intrinsic configurations are 
already parity eigenstates with eigenvalue $\pi=+1$. For the same 
reason, the  negative parity $\pi=-1$ projected wave function only 
makes sense along the $Q_{30}$=0 axis when a limiting procedure is 
considered. The evaluation of physical quantities in this case is 
subject to numerical inaccuracies consequence of evaluating the 
ratio of two small quantities (the denominator is the norm of the 
projected negative parity state that is zero in this case) and 
alternative expressions, obtained by considering explicitly the 
$Q_{30}$= 0 limit \cite{egi91}, are required for a sound numerical 
evaluation of those quantities.  Note however (see, Fig. \ref
{evol-min-gamma-Sm}) that the negative parity projected energy 
increases rapidly while approaching the $Q_{30}$=0 configuration and 
therefore it does not play a significant role in the subsequent 
discussion of the corresponding PPPESs. As a consequence, we have 
omitted this quantity along the $Q_{30}$=0 axis.

As in the mean field case, the results with D1S and D1M show a
striking similarity and therefore only the D1S results will be
discussed. The comparison between the  MFPESs  in Fig. \ref{fig_mfq2q3_Sm} and the 
PPPESs in Fig. \ref{fig_ppq2q3_Sm}, clearly illustrates the 
topological changes induced by the restoration of the reflection 
symmetry. In general, the quadrupole moments $Q_{20}$ corresponding 
to the  absolute minima of the PPPESs, remain quite close to the 
ones obtained at the HFB level (see, Tables I and  II) increasing 
their values as more  neutrons are added  for each of the Sm and Gd 
chains. On the other hand, the situation is quite  different along 
the $Q_{30}$ direction. To obtain a more quantitative understanding 
of the evolution of the PPPESs, we have plotted in Fig. \ref{evol-min-gamma-Sm}
  the parity-projected energy curves for selfconsistent
$Q_{20}$ values, as a 
function of the octupole moment $Q_{30}$ for the nucleus $^{150}$ 
Sm. The corresponding HFB energy curves are also included for 
comparison.  For $^{150}$Sm, and all the other nuclei considered in 
the present study, the negative parity curves always show a well 
developed minimum at $Q_{30}$ values in the range 
1.50-1.75 b$^{3/2}$. On the other hand, and regardless of the particular 
version of the Gogny-EDF employed, the $\pi=+1$ curves always 
display a characteristic pocket \cite{butler_96,bon88,egi91} with a 
minimum at $Q_{30}$ = 0.50-0.75 b$^{3/2}$. In the spirit of the 
Variation after Projection procedure, 
the configuration yielding the minimum of the positive (negative) parity 
projected energy as a function of $Q_{20}$ and $Q_{30}$ is to be 
associated with the positive (negative) parity state. As a 
consequence of this "minimization after projection" the intrinsic states
for each parity have different deformations. The positive parity ground 
state gains an amount of energy $E_{corr}^\textrm{par proj}$ given by 
\begin{equation} \label{formula-correner-pp} 
E_{corr}^\textrm{par proj} =  E_{HFB}^{g.s.} - E_{\pi=+1}^{g.s.} 
\end{equation} 
where, $E_{\pi=+1}^{g.s.}$ corresponds to the absolute 
minima of the positive parity PPPESs and $E_{HFB}^{g.s.}$ to the HFB 
ground state energies, i.e., the absolute minima of the MFPESs.  
Regardless of the Gogny-EDF employed, they are always smaller than 900 keV in 
each of  the considered  nuclei. This correlation energy has to be 
compared to the correlation energy gained by configuration mixing 
(see also Fig. \ref{Correlation_Ener_GCM} below).

%
%

\section{Generator Coordinate Method.}
 \label{BYMF-RESULTS-GCM}

According to the discussions in previous sections, it can be
concluded that not only the plain HFB results of Sec. \ref{MF-RESULTS}, 
but even the parity projection ones, may not be 
sufficient to decide  whether, as suggested in Ref. \cite{zhang_10}, 
there exists a transition to an octupole deformed regime in the 
considered nuclei in addition to the transitional behavior along the 
$Q_{20}$-direction \cite{ours_x5_1,ours_x5_2}. Within this context,  
$(Q_{20},Q_{30})$-GCM calculations are needed in order to verify the 
stability of the quadrupole and/or octupole deformation effects 
encountered in both the MFPESs and the PPPESs for the considered Sm 
and Gd nuclei. One should also keep in mind that in the framework of 
such a dynamical 2D-GCM treatment, not only the mean field energy surface  but also 
the underlying collective inertia  plays a role. 

\begin{figure*}
\includegraphics[width=0.98\textwidth]{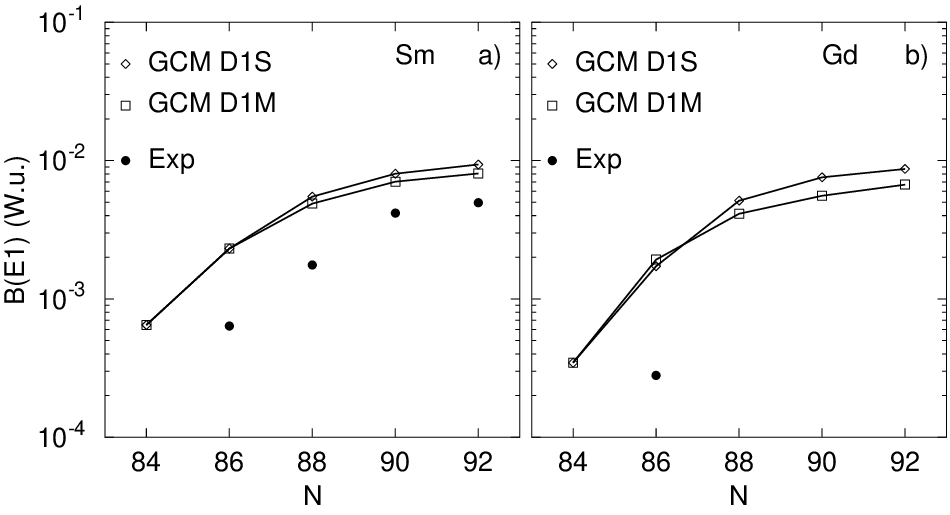}%

\includegraphics[width=0.98\textwidth]{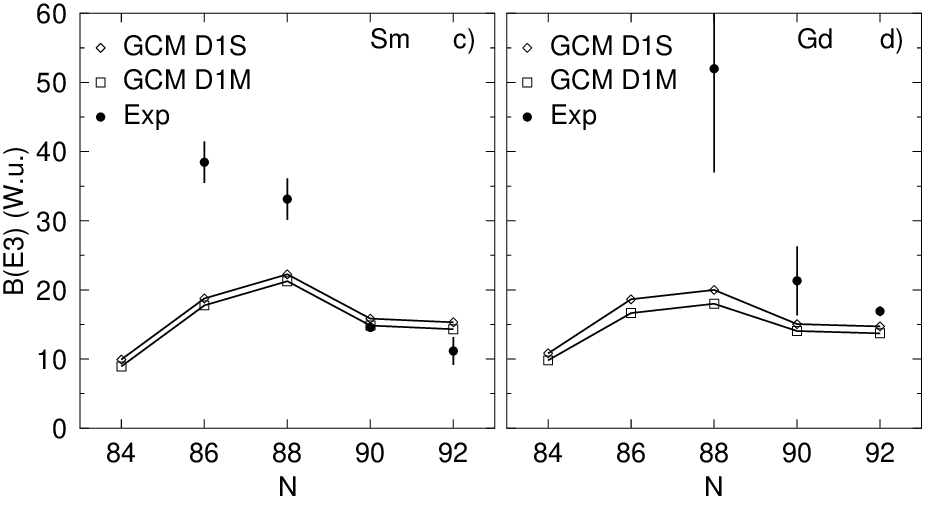}%
\caption{Theoretical and experimental transition rates 
$B(E1,1^{-} \rightarrow 0^{+})$ (panels a and b) and $B(E3,3^{-} \rightarrow 0^{+})$ (panels c and d)
for the nuclei $^{146-154}$Sm and $^{148-156}$Gd. Results are shown for the Gogny-EDFs D1S and D1M.
Experimental results for $B(E1,1^{-} \rightarrow 0^{+})$ rates are extracted from 
Ref. \cite{butler_96} while the experimental $B(E3,3^{-} \rightarrow 0^{+})$ rates are taken from Ref. \cite{Kibedi}.
}
\label{BE3_GCM}
\end{figure*}

The superposition of HFB states 
\begin{equation} \label{GCM-WF}
| {\Psi}_{\sigma}^{\pi} \rangle = \int d{\bf Q} f_{\sigma}^{\pi} ({\bf Q}) | {\Phi} ({\bf Q}) \rangle
\end{equation}
is used to define the GCM wave functions $| {\Psi}_{\sigma}^{\pi} \rangle$.
In the integration domain both positive and negative 
octupole moments  $Q_{30}$ are included. 
The GCM amplitudes $f_{\sigma}^{\pi} ({\bf Q})$ are the solutions of the 
Hill-Wheeler (HW) equation \cite{rs}
\begin{equation} \label{HW-equation}
\int d{\bf Q}^{'} 
\left(
{\cal{H}}({\bf Q},{\bf Q}^{'}) - E_{\sigma}^{\pi}
{\cal{N}}({\bf Q}, {\bf Q}^{'})
\right)
f_{\sigma}^{\pi} ({\bf Q}^{'}) = 0.
\end{equation}
The GCM hamiltonian  ${\cal{H}}({\bf Q}, {\bf Q}^{'})$
and norm ${\cal{N}}({\bf Q}, {\bf Q}^{'})$ kernels are  given by
\begin{eqnarray} \label{GCM-PROJEDF-hnk}
{\cal{H}}({\bf Q}, {\bf Q}^{'}) &=&
\langle {\Phi} ({\bf Q}) | 
\hat{H} [\rho^{GCM}(\vec{r}) ] | {\Phi} ({\bf Q}^{'}) \rangle,
\nonumber\\
{\cal{N}}({\bf Q}, {\bf Q}^{'}) &=&
\langle {\Phi} ({\bf Q}) |  {\Phi} ({\bf Q}^{'}) \rangle
\end{eqnarray}
where in the evaluation of  ${\cal{H}}({\bf Q}, {\bf Q}^{'})$  the 
{\it{mixed density}} prescription is used
\begin{equation}
\rho^{GCM}(\vec{r})= 
\frac{
\langle {\Phi} ({\bf Q}) | \hat{\rho}({\vec{r})} | {\Phi} ({\bf Q}^{'}) \rangle
}
{
\langle {\Phi} ({\bf Q}) | {\Phi} ({\bf Q}^{'}) \rangle
}.
\end{equation}

As in the parity projection case the Hamiltonian kernel ${\cal{H}}({\bf Q}, 
{\bf Q}^{'})$ is also supplemented with first order corrections in 
both proton and neutron numbers \cite{har82,bon90,egi91}.

The solution of the HW equation (\ref{HW-equation}) provides the 
energies $E_{\sigma}^{\pi}$ corresponding to the ground ($\sigma=1$) 
and excited  ($\sigma=2,3, \dots$) states. The parity of each of 
these states is given by the behavior of $f_{\sigma}^{\pi} ({\bf Q})$
under the $Q_{30} \rightarrow -Q_{30}$ exchange. This is a 
consequence of the invariance under reflection symmetry of the GCM 
Hamiltonian kernels. For details on the solution of Eq. (\ref
{HW-equation}), the reader is referred, for example, to Refs. \cite
{rs,rod02,bon90}.  Since the $| {\Phi}({\bf Q}) \rangle$  basis states are 
not orthogonal, the functions  $f_{\sigma}^{\pi} ({\bf Q})$ of  
Eq. (\ref{GCM-WF}) can not be interpreted as probability 
amplitudes. One then introduces (see, for example, Refs. \cite
{rs,rod02}) the collective wave functions 
\begin{equation} \label{cll-wfs-HW} 
G_{\sigma}^{\pi} ({\bf Q}) =   \int d{\bf Q}^{'} {\cal
{N}}^{\frac{1}{2}}({\bf Q}, {\bf Q}^{'})  f_{\sigma}^{\pi}({\bf 
Q}^{'}),
\end{equation} 
which are orthogonal and therefore their 
modulus squared $|G_{\sigma}^{\pi} ({\bf Q})|^{2}$ has the meaning 
of a probability amplitude. It is easy to show that the parity of 
the collective wave functions $G_{\sigma} (Q_{20}, Q_{30})$ under the exchange 
$Q_{30} \rightarrow -Q_{30}$ corresponds to the spatial parity 
operation in the correlated wave functions built up from them. The 
inclusion of octupole correlations immediately restores the 
reflection symmetry spontaneously broken at the mean field level and
grants the use of a parity label $\pi$ for the GCM quantities. 

The collective wave functions of Eq. (\ref{cll-wfs-HW}) can be used to
express overlaps of operators between GCM wave functions
in a more convenient way
\begin{equation} \label{OOverGCM}
\langle \Psi_{\sigma}^{\pi} | \hat{O} |\Psi_{\sigma '}^{\pi '} \rangle =
\int d {\bf Q} d {\bf Q}^{'} G_{\sigma}^{\pi \,*} ({\bf Q}) {\cal O } ({\bf Q}, {\bf Q}^{'})
G_{\sigma '}^{\pi '} ({\bf Q}^{'}),
\end{equation}
with the kernels
\begin{equation} \label{Okernel}
{\cal{O}} ({\bf Q}, {\bf Q}^{'})  = \int d{\bf Q}^{''} d{\bf Q}^{'''}
{\cal{N}}^{-\frac{1}{2}}({\bf Q}; {\bf Q}^{''}) \langle {\bf Q}^{''} | \hat{O}
| {\bf Q}^{'''} \rangle {\cal{N}}^{-\frac{1}{2}}({\bf Q}^{'''}; {\bf Q}^{'})
\end{equation}
given in terms of the operational square root of the overlap kernel that
is defined by the property
\begin{equation} \label{Oversqrt}
{\cal{N}}({\bf Q}; {\bf Q}^{'})  = \int d{\bf Q}^{''} 
{\cal{N}}^{\frac{1}{2}}({\bf Q}; {\bf Q}^{''}) {\cal{N}}^{\frac{1}{2}}({\bf Q}^{''}; {\bf Q}^{'}).
\end{equation}

\begin{figure*}
\includegraphics[width=0.82\textwidth]{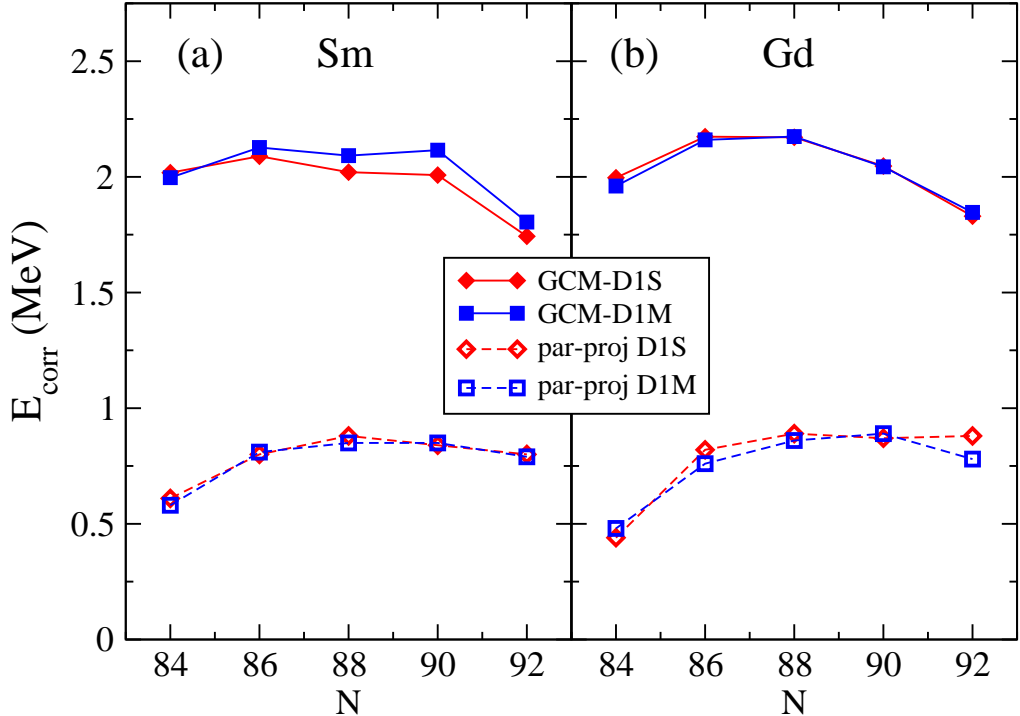}
\caption{(Color online) The 2D-GCM correlation energies for the nuclei 
$^{146-154}$Sm (panel a) and $^{148-156}$Gd (panel b) are 
shown, as functions of the neutron number. The correlation energies 
stemming from the restoration of reflection
symmetry are also included. Results are shown for both Gogny-D1S and Gogny-D1M
EDFs. For more details, see main text.
}
\label{Correlation_Ener_GCM}
\end{figure*}

The solution of Eq. (\ref{HW-equation}) allows the calculation of 
physical observables like the energy splitting between positive and 
negative parity states as well as B(E1) and B(E3) transition 
probabilities. In the present study time reversal symmetry is 
preserved and therefore only excited states with an average angular 
momentum zero can be accounted for. Genuine $1^{-}$ and $3^{-}$ 
states, on the other hand, will require to consider cranking HFB 
states \cite{rs,gar98}, a calculation which is out of the scope of the present 
work. We assume here that the cranking rotational energy of the 
$1^{-}$ and $3^{-}$ states is much smaller than the excitation 
energy of the negative-parity bandhead and therefore it can be neglected. 
For the reduced transition probabilities $B(E1,1^{-} \rightarrow 0^{+} )$ and $B(E3,3^{-} 
\rightarrow 0^{+} )$  the rotational model approximation for K=0 
bands has been used 
\begin{equation} \label{exp-BE1}
	B(E\lambda,\lambda^{-} \rightarrow 0^{+} ) = \frac{e^{2}}{4 \pi} 
	\Big{|} \langle \Psi_{\sigma}^{\pi=-1} | \hat{{\cal O}}_{\lambda} |\Psi_{\sigma=1}^{\pi=+1} \rangle \Big{|}^{2}
\end{equation}
where $\sigma$ corresponds to the first GCM excited state of negative 
parity. The electromagnetic transition operators  $\hat{\cal{O}}_{1}$ and $\hat{\cal{O}}_{3}$ represent the 
dipole moment operator of Eq. (\ref{dipole-operator}) and the proton 
component $\hat{Q}_{30, \textrm{prot}}$ of the octupole operator, respectively. 
The evaluation of the overlap is carried out using Eq. (\ref{OOverGCM}).

In Fig. \ref{coll-wfs-GCMQ2Q3-Sm}  the collective 
probability amplitude $|G_{\sigma}^{\pi} (Q_{20},Q_{30}) |^{2}$ of Eq. (\ref
{cll-wfs-HW}), obtained from the solution of the HW equation (\ref
{HW-equation}) are plotted. As a typical example, results for the 
$^{146,150,154}$Sm isotopes and the Gogny-D1S EDF are presented. For 
other nuclei and Gogny parametrizations, the results look very 
similar. The left panels in Fig. \ref{coll-wfs-GCMQ2Q3-Sm} 
correspond to the ground state wave functions  (i.e., $\sigma=1$ and 
$\pi$=+1) while the right panels correspond to the  
lowest-lying $\pi$=-1 states  $\sigma=3$ for $^{146}$Sm and $\sigma=2$ for the others.

\begin{figure*}
\includegraphics[width=0.85\textwidth]{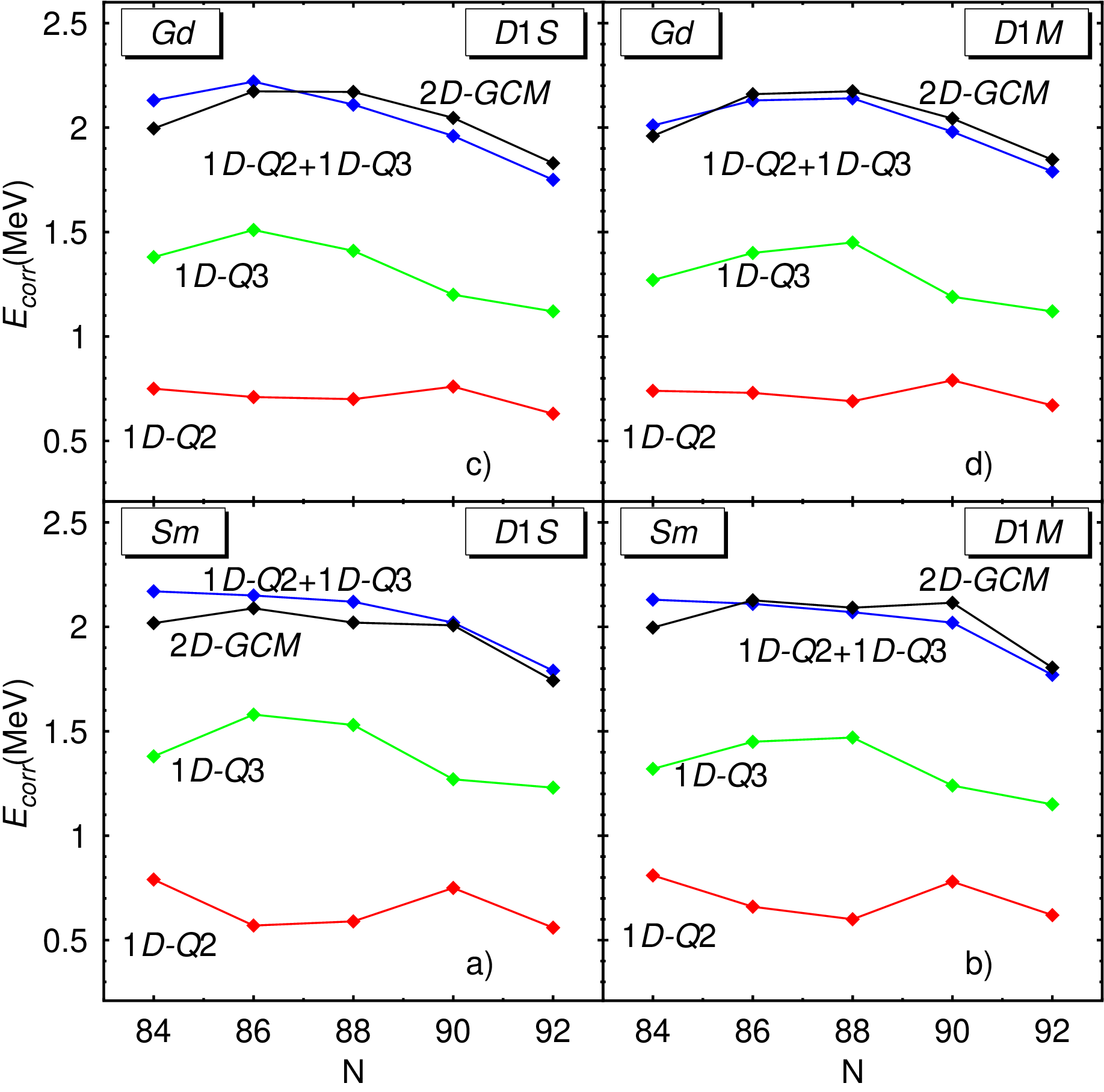}
\caption{(Color online) The sum 
(blue curve named 1D-Q2+1D-Q3)
of the correlation energies 
$E_{corr}^{1D-Q3}$
[Eq.(\ref{1D-OCT-GCM-corr-ener})]
(green curve named 1D-Q3)
and 
$E_{corr}^{1D-Q2}$
[Eq.(\ref{1D-RS-GCM-corr-ener})]
(red curve named 1D-Q2)
is compared with the correlation energy 
$E_{corr}^{2D-GCM}$
[Eq.(\ref{2D-GCM-correlation-energies})]
(black curve named 2D-GCM)
provided by the full 2D-GCM calculations
in $^{146-154}$Sm and $^{148-156}$Gd.
Results are shown for both Gogny-D1S and Gogny-D1M
EDFs. For more details, see main text.
}
\label{compare-2D-1D-GCM}
\end{figure*}

The ground state collective  probability amplitude $|G_{\sigma=1}^{\pi=+1} 
(Q_{20},Q_{30}) |^{2}$ reach a global maximum  at $Q_{30}=0$ 
pointing  to the octupole-soft character of the ground states in 
$^{146-154}$Sm. The spreading along the
octupole direction is large for $^{150, 154}$Sm indicating octupole
softness in these nuclei. For the negative parity collective wave functions
the maximum is always located at a non zero value of $Q_{30}$ as could be
anticipated from the parity projection results. For $^{150, 154}$Sm
the wave function spreads out farther along $Q_{30}$ than in previous cases 
in agreement with the octupole softness of their ground states.

To have a more quantitative characterization of the collective
wave functions we have computed mean values of relevant operators
(see Eq. (\ref{OOverGCM})). The first is the average of the
quadrupole moment defined as
\begin{equation}
	(\bar{Q}_{20})_{\sigma}^{\pi}= 
	\langle {\Psi}_{\sigma}^{\pi} | \hat{Q}_{20}| {\Psi}_{\sigma}^{\pi} \rangle.
\end{equation}
For negative parity operators like the octupole or the dipole moment
the above averages are zero by construction and therefore a meaningful  
averaged quantity has to be defined as 
\begin{equation}
	\bar{O}_{\sigma}^{\pi}= 4 \int_{Q_{30}>0, Q_{30}^{'}>0} d {\bf Q} d {\bf Q'} G_{\sigma}^{\pi \,*} ( {\bf Q} )  {\cal O } ({\bf Q}, {\bf Q}^{'})
G_{\sigma}^{\pi} ({\bf Q}^{'})
\end{equation}
where a restriction to positive values of the octupole moment has been made.
The average quadrupole $\bar{Q}_{20}^{(+)}$ and octupole $\bar{Q}_{30}^{(+)}$
moments for the ground state ($\sigma =1$) are listed in  Tables III and IV. 
The $\bar{Q}_{20}^{(+)}$ moments follow a trend similar 
to the one found within the HFB approximation increasing their 
values as more neutrons are added in a given isotopic chain. On the 
other hand, the isotopic trend predicted for 
$\bar{Q}_{30}^{(+)}$ is quite different to the one predicted 
at the mean field level. As discussed in Sec. \ref{MF-RESULTS}, at 
the Gogny-HFB level only the N=86 and 88 isotones $^{148,150}$Sm and 
$^{150,152}$Gd display non vanishing (static) octupole moments (see, 
Tables I and II). Nevertheless, after both projection onto $\pi=+1$ 
and dynamical $(Q_{20},Q_{30})$-fluctuations are considered at the 
2D-GCM level, the octupole deformation effects predicted for 
$^{148,150}$Sm and $^{150,152}$Gd are reduced to more than half of 
their mean field values. At variance with the HFB results, the 
nuclei $^{146,152,154}$Sm and $^{148,154,156}$Gd exhibit dynamical 
ground state octupole moments $\bar{Q}_{30}^{(+)} 
\approx$ 0.40-0.50 b$^{3/2}$. We conclude 
that, regardless of the particular version of the Gogny-EDF 
employed, our 2D-GCM calculations suggest a dynamical shape/phase 
transition  from weakly ($^{146}$Sm and $^{148}$Gd) to well 
quadrupole  deformed ($^{154}$Sm and $^{156}$Gd) ground states as 
well as a transition to an octupole vibrational regime in  the 
considered Sm and Gd nuclei.

For the lowest-lying negative parity states, the   dynamical 
octupole $\bar{Q}_{30}^{(-)}$ and quadrupole 
$\bar{Q}_{20}^{(-)}$ moments, computed with the 
corresponding 2D-GCM states $| {\Psi}_{\sigma=2}^{\pi=-1} \rangle$ or 
$| {\Psi}_{\sigma=3}^{\pi=-1} \rangle$, are also 
listed in Tables III and IV. It should be noted that the largest 
values of the octupole deformations $\bar{Q}_{30}^{(+)}$ and 
$\bar{Q}_{30}^{(-)}$ always correspond to the N=88 isotones 
$^{150}$Sm and $^{152}$Gd.

The values of the ground state  dipole moments 
$\bar{D}_{0}^{(+)}$ are less predictable 
than the averages of the quadrupole and octupole moments discussed 
previously as the behavior of $D_{0}$ for the HFB states strongly 
depends upon the orbitals occupied and those change rapidly with 
deformation. The comparison of the dipole moments with available 
experimental data \cite{butler_96} is presented in panels a) and b) 
of Fig. \ref{Dipole_GCM}. In particular, the comparison between  the 
HFB results (see, Tables I and II), $\bar{D}_{0}^{(+)}$ and 
experimental values  clearly reveal the limitations of the HFB 
approximation to predict dipole moments in this region of the 
nuclear chart.

Another physical observable is the energy splitting between the 
lowest lying $\pi=+1$ and $\pi=-1$ states. The results for $^{146-154}$
Sm and $^{148-156}$Gd are compared in Fig. \ref{splitting01_GCM} 
with available experimental $0^{+}-1^{-}$ and $0^{+}-3^{-}$ energy
splittings \cite{exp_ensdf}. As already mentioned, in the present 
study we are not able to account for genuine $1^{-}$ and/or $3^{-}$ 
states that require, for example, the use of cranking HFB  states 
\cite{rs,gar98}. With this in mind and regardless of the 
Gogny-EDF employed, a reasonable agreement between the theoretical 
and experimental energy splittings is observed. The remaining discrepancies 
imply that correlations other than (axial) quadrupole-octupole fluctuations 
could also be required. In particular, the time-odd components of the 
Gogny-EDF, incorporated throughout cranking calculations, should be 
further investigated within the present 2D-GCM framework. Let us 
mention that the results are compatible with the ones obtained in Ref. \cite
{egi92} using a one-dimensional collective Hamiltonian whose parameters 
are derived from octupole constrained calculations. This is also
the case with the systematic calculations of Ref. \cite{rob11b} using a
GCM with the octupole degree of freedom as generating coordinate.

In panels a) and b) of Fig. \ref{BE3_GCM}, the $B(E1,1^{-} 
\rightarrow 0^{+} )$ reduced transition probabilities of Eq. (\ref{exp-BE1}) 
are compared with experimental data \cite{butler_96}. It is very 
satisfying to observe how, without resorting to any effective 
charges, the predicted $B(E1,1^{-} \rightarrow 0^{+} )$ values in Sm 
nuclei follow the experimental isotopic trend with a slight 
improvement in the case of the Gogny-D1M EDF. In panels c) and d) of the same 
figure, we compare the $B(E3,3^{-} \rightarrow 0^{+} )$ transition 
rates of Eq. (\ref{exp-BE1})] with available data \cite{Kibedi}. The
predicted $B(E3,3^{-} \rightarrow 0^{+} )$ values 
reproduce quite well the experimental ones in the case of 
$^{152,154}$Sm and $^{154,156}$Gd. On the other hand, from the 
comparison between ours and the $B(E1,1^{-} \rightarrow 0^{+} )$ and 
$B(E3,3^{-} \rightarrow 0^{+} )$ rates obtained in Refs. \cite
{egi92} and \cite{rob11b}, we can conclude that they are, 
to a large extent, not very sensitive to quadrupole fluctuations. 

In panels a) and b) of Fig. \ref{Correlation_Ener_GCM}, the correlation 
energies defined as the difference between the reference HFB ground state
energy and the 2D-GCM one 
\begin{equation} \label{2D-GCM-correlation-energies}
E_{corr}^{2D-GCM} = E_{HFB}^{g.s} - E_{\sigma=1}^{\pi=+1}
\end{equation} 
are ploted. The parity restoration correlation 
energies $E_{corr}^\textrm{par proj}$ of Eq. (\ref{formula-correner-pp}) 
are also included for comparison. The predicted isotopic trends and quantitative 
values of $E_{corr}^{2D-GCM}$ are quite similar for both Gogny-D1S 
and Gogny-D1M EDFs. The correlation energies $E_{corr}^{2D-GCM}$ exhibit a 
relatively weak dependence with  neutron number with values oscillating 
between 1.74  and 2.09 MeV  for Sm and between 1.83 and 2.17 MeV for Gd nuclei. 
The smooth variation of the correlation energy is, however, of the same order
of magnitude as the rms for the binding energy in modern nuclear mass tables \cite{gogny-d1m}
and therefore the dynamical octupole correlation energies should be considered 
in improved versions of them.

A rough estimate of the contribution of the $(Q_{20},Q_{30})$ 
fluctuations to the correlation energies can be obtained by subtracting
to the total correlation energy the parity projected one. Those contributions
range in between 0.94 and 1.41 MeV for Sm isotopes and in between 0.94
and 1.56 MeV for Gd isotopes. The oscillations are slightly larger than for
the total correlation energy.

In order to determine the contributions of each degree of freedom in the
results obtained we have also performed one dimensional GCM calculations
along each of the degrees of freedom. First, the octupole moment has been
used as generating coordinate. For each octupole moment considered the quadrupole moment 
corresponds to the minimum energy. The octupole moments of the
generating wave functions are taken in the range -7 b$^{3/2}$ $\le Q_{30} \le$ 7 
b$^{3/2}$ and with a mesh size $\delta Q_{30}$= 0.25 b$^{3/2}$. The
1D-GCM ansatz is
\begin{equation} \label{GCM-WF-1D-Q3}
| {\Psi}_{\sigma, 1D-Q3}^{\pi} \rangle =
 \int dQ_{30} f_{\sigma, 1D-Q3}^{\pi} (Q_{30}) | {\Phi} (Q_{30}) \rangle
\end{equation}
given in terms of the HFB states $| {\Phi} (Q_{30}) \rangle$. Note that no 
quadrupole constraint is imposed in these calculations. From the 
1D-Q3 ground  state energies  we can 
computed the 1D octupole correlation energy 
\begin{equation} \label{1D-OCT-GCM-corr-ener}
E_{corr}^{1D-Q3} = E_{HFB}^{g.s} - E_{\sigma=1,1D-Q3}^{\pi = +1} 
\end{equation}  
This quantity is displayed in panels a) to d) of Fig. \ref
{compare-2D-1D-GCM} for the considered Sm and Gd nuclei. It has to
be mentioned that this type of calculations have been carried out
for all possible even-even nuclei with several parametrizations of the
Gogny force in \cite{rob11b}.

In a second step, GCM calculations with the quadrupole degree of 
freedom ($Q_{30}=0$, i.e., reflection symmetry is preserved) as 
generating coordinate have been performed. The $Q_{20}$ values used 
are in the interval -30b $\le Q_{20} \le$ 30b with $\delta Q_{20}$= 
0.6 b. The GCM wave functions 
\begin{equation} \label{GCM-WF-1D-Q2}
| {\Psi}_{\sigma, 1D-Q2} \rangle =
 \int dQ_{20} f_{\sigma, 1D-Q2} (Q_{20}) | {\Phi} (Q_{20}) \rangle
\end{equation}
are defined in terms of the  states  $| {\Phi} (Q_{20}) \rangle$.
The corresponding correlation energy 
\begin{equation} \label{1D-RS-GCM-corr-ener}
E_{corr}^{1D-Q2} =  E_{HFB}^{g.s}-E_{\sigma=1,1D-Q2} 
\end{equation}
is displayed in panels a) to d) of Fig. \ref{compare-2D-1D-GCM}. 

In panels a) to d) of Fig. \ref{compare-2D-1D-GCM}, we compare the 
sum $E_{corr}^{1D-Q2+1D-Q3} =E_{corr}^{1D-Q3} + E_{corr}^{1D-Q2}$ 
with the correlation energies $E_{corr}^{2D-GCM}$ of Eq. (\ref{2D-GCM-correlation-energies}). 
For the particular 
set of Sm and Gd nuclei considered in the present study and 
regardless of the Gogny-EDF employed, the correlation energies 
provided  by the full 2D-GCM are very well reproduced by the sum of 
the ones obtained in the framework of the 1D-GCM approximations  
(\ref{GCM-WF-1D-Q3}) and (\ref{GCM-WF-1D-Q2}). Obviously, this is far 
from being a general statement and further explorations in other 
regions of the nuclear chart, specially those showing shape coexistence
already at $Q_{30}=0$ are required. Nevertheless, the kind of decoupling observed 
in our results may be potentially relevant to incorporate 
correlation energies stemming from parity restoration and octupole 
fluctuations in large scale calculations of nuclear masses based on 
the Gogny-EDF (see, for example, \cite{delaroche10}) as well as in 
future fitting protocols beyond the most recent D1M parametrization 
\cite{gogny-d1m} of the Gogny EDF.

\section{Conclusions}\label{CONCLU}

Calculations have been carried out using the GCM method and with the multipole moments 
$Q_{20}$ and $Q_{30}$ as generating coordinates  for
several Sm and Gd isotopes and with different parametrizations of the
Gogny force. The results from different parametrizations are very close to each other
indicating again that the D1M parametrization of the Gogny force performs as well as D1S in spectroscopic
calculation. The comparison with experimental data is fairly good both
for excitation energies and electromagnetic transition probabilities 
reassuring the predictive power of the Gogny class of EDFs. Comparison of
the 2D GCM results with the outcome of previous 1D Collective Schrodinger equation calculations in
the same region points out to a decoupling of the dynamics of the 
quadrupole and octupole degrees of freedom. This conclusion is reinforced
by the comparison of the 2D correlation energies with the sum of correlation
energies along each of the degrees of freedom. Correlation energies show
a smooth behavior with neutron number with differences between different isotopes
as large as 200 keV. Although these differences are small, they can be 
relevant for theories aiming at providing accurate mass tables for
applications requiring accurate reaction rates that depend on their 
energetic balance.

\begin{acknowledgments}

Work supported in part by MINECO (Spain) under 
grants Nos. FPA2009-08958, FIS2009-07277 and FIS2011-23565 and the 
Consolider-Ingenio 2010 Programs CPAN CSD2007-00042 and MULTIDARK 
CSD2009-00064. 

\end{acknowledgments}

\end{document}